\documentclass[aps,prl,twocolumn,superscriptaddress,showpacs]{revtex4-1}

\usepackage{graphicx}
\usepackage{dcolumn}
\usepackage{bm}
\usepackage{color}
\usepackage[colorlinks,bookmarks=false,citecolor=blue,linkcolor=red,urlcolor=blue]{hyperref}

\begin{document}

\title{Magnetization of SrCu$_2$(BO$_3$)$_2$ in ultrahigh magnetic fields up to 118~T}

\author{Y.~H.~Matsuda}
\email{ymatsuda@issp.u-tokyo.ac.jp}
\affiliation{Institute for Solid State Physics, University of Tokyo, Kashiwa, Chiba 277-8581, Japan}

\author{N.~Abe}
\affiliation{Institute for Solid State Physics, University of Tokyo, Kashiwa, Chiba 277-8581, Japan}

\author{S.~Takeyama}
\affiliation{Institute for Solid State Physics, University of Tokyo, Kashiwa, Chiba 277-8581, Japan}

\author{H.~Kageyama}
\affiliation{Graduate School of Engineering, Kyoto University, Nishikyouku, Kyoto 615-8510, Japan}

\author{P.~Corboz}
\affiliation{Theoretische Physik, ETH Z\"urich, CH-8093 Z\"urich, Switzerland}

\author{A.~Honecker}
\affiliation{Institut f\"ur Theoretische Physik, Georg-August-Universit\"at G\"ottingen,
     D-37077 G\"ottingen, Germany}
\affiliation{Fakult\"at f\"ur Mathematik und Informatik, Georg-August-Universit\"at G\"ottingen,
     D-37073 G\"ottingen, Germany}

\author{S.~R.~Manmana}
\affiliation{Institut f\"ur Theoretische Physik, Georg-August-Universit\"at G\"ottingen,
     D-37077 G\"ottingen, Germany}

\author{G.~R.~Foltin}
\affiliation{Lehrstuhl f\"{u}r Theoretische Physik I, Otto-Hahn-Stra\ss e 4, TU Dortmund, 44221 Dortmund, Germany}

\author{K.~P.~Schmidt}
\affiliation{Lehrstuhl f\"{u}r Theoretische Physik I, Otto-Hahn-Stra\ss e 4, TU Dortmund, 44221 Dortmund, Germany}

\author{F.~Mila}
\affiliation{Institute of Theoretical Physics, Ecole Polytechnique F\'ed\'erale de Lausanne (EPFL), 1015 Lausanne, Switzerland}

\date{August 19, 2013}
\begin{abstract}

The magnetization process of the orthogonal-dimer antiferromagnet 
SrCu$_2$(BO$_3$)$_2$ is investigated in high magnetic fields of up to 
118~T. A 1/2 plateau is clearly observed in the field range 84 to 108~T 
in addition to 1/8, 1/4 and 1/3 plateaux at lower fields. Using a 
combination of state-of-the-art numerical simulations, the main features 
of the high-field magnetization, a 1/2 plateau of width 24~T, a 1/3 
plateau of width 34~T, and no 2/5 plateau, are shown to agree 
quantitatively with the Shastry-Sutherland model if the ratio of inter- 
to intra-dimer exchange interactions $J^{\prime}/ J =0.63$. It is further 
predicted that the intermediate phase between the 1/3 and 1/2 plateau is 
not uniform but consists of a 1/3 supersolid followed by a 2/5 supersolid 
and possibly a domain-wall phase, with a reentrance into the 1/3 
supersolid above the 1/2 plateau.

\end{abstract}
\pacs{
75.10.Jm;	
75.60.Ej; 	
75.40.Mg	
}
\maketitle

Geometrical frustration can induce very interesting phases in quantum magnets \cite{HFM}.
For instance, the orthogonal dimer antiferromagnet SrCu$_2$(BO$_3$)$_2$ exhibits 
fascinating phenomena due to frustration.
The nearest neighbor (NN) $S$=1/2 spins of Cu ions are antiferromagnetically coupled and form singlet dimers
through the exchange interaction $J$.
Since the inter-dimer exchange interaction $J^{\prime}$ between the next nearest-neighbor (NNN) Cu ions 
is antiferromagnetic as well, the orthogonal configuration induces geometrical frustration
\cite{Kage99}.
Quite remarkably, the crystal lattice is topologically equivalent to the Shastry-Sutherland lattice 
that was initially investigated out of pure theoretical interest \cite{SS}. 
Since its discovery, SrCu$_2$(BO$_3$)$_2$ has thus logically been the subject of a vast number of experimental and 
theoretical studies~\cite{Miyahara03JPCM,Takigawa10,TM}.
 
Quantum phase transitions have been theoretically predicted to take place when the ratio 
$J^{\prime} /J$ is tuned. It is clear that the ground state is a product of dimer singlets if $J^{\prime} /J = 0$, 
and that it supports antiferromagnetic N\'eel order when $J^{\prime} /J \rightarrow +\infty$. 
An intermediate gapped plaquette phase has been predicted to appear \cite{Koga2000,Takushima01,Chung01,Laeuchli02}  when $0.675 \lesssim
J^{\prime} /J \lesssim 0.77$ \cite{lou12,corboz2013}. SrCu$_2$(BO$_3$)$_2$ is believed to be located at $J^{\prime} /J \simeq 0.63$, thus to have an exact dimer singlet ground state \cite{Miyahara03JPCM,Takigawa10}.

In addition to the interest raised by the exotic ground state of the Shastry-Sutherland model, the presence of several magnetization plateaux 
in SrCu$_2$(BO$_3$)$_2$ has attracted significant attention. 
Distinct 1/8, 1/4, and 1/3 plateaux have been reported early on in the
magnetization process \cite{Kage99,Onizuka00}.
More recently, additional
plateaux between 1/8 and 1/4 have been observed \cite{Sebastian08,Takigawa12}, and evidence in favor of the presence of 
the long predicted 1/2 plateau has been provided by magnetostriction measurements \cite{Jaime12}. 
However, the entire 1/2 plateau phase has not been unveiled in Ref.~\onlinecite{Jaime12} because of the technical upper limit of the magnetic field at 100.75~T. 

The 1/2 plateau has been predicted to be less stable than the 1/3 plateau and to disappear for
large $J'/J$ \cite{MT00PRB}.
In fact, according to Ref.~\onlinecite{Miyahara00PRB}, the length of the 1/2 plateau is less than half that 
of the 1/3 plateau, although the 1/2 plateau can be expected to be quite stable considering the checkerboard 
pattern of the triplet excitation suggested by the boson picture.
Hence, the experimental determination of the stability range of the 1/2 plateau is of particular interest in itself, and also 
important for checking the validity of the theoretical model. 
Moreover, in addition to the 1/2 plateau, exotic high-field spin states have been predicted 
such as supersolid phases between the 1/3 and 1/2 plateaux and above the 1/2 plateau
\cite{MT00PRB,lou12}.
The quantum spin state realized when the density of triplets becomes very high has not been uncovered yet. 

In the present work, we have investigated the spin states of SrCu$_2$(BO$_3$)$_2$ by magnetization measurements 
in high magnetic fields up to 118~T.
A clear 1/2 magnetization plateau phase has been observed in the field range from 84 to 108~T, and at the upper critical field, 
a sharp magnetization increase suggests a first-order phase transition.
Theoretical calculations based on the infinite projected entangled-pair state (iPEPS) tensor network algorithm \cite{verstraete2004,jordan2008,corboz2010,corboz2011,bauer2011}, exact diagonalizations,
density-matrix renormalization group (DMRG) simulations and series expansions have shown that the 1/2 and 1/3 plateaux
can be quantitatively reproduced by the Shastry-Sutherland model with a ratio $J'/J\simeq 0.63$, and they predict a variety of
exotic phases between the 1/3 and 1/2 plateaux and above the 1/2 plateau, including several types of supersolid phases, in particular
a first-order transition to a 1/3 supersolid above the upper critical field of the 1/2 plateau.

\begin{figure}[b]
\includegraphics[width=1\columnwidth]{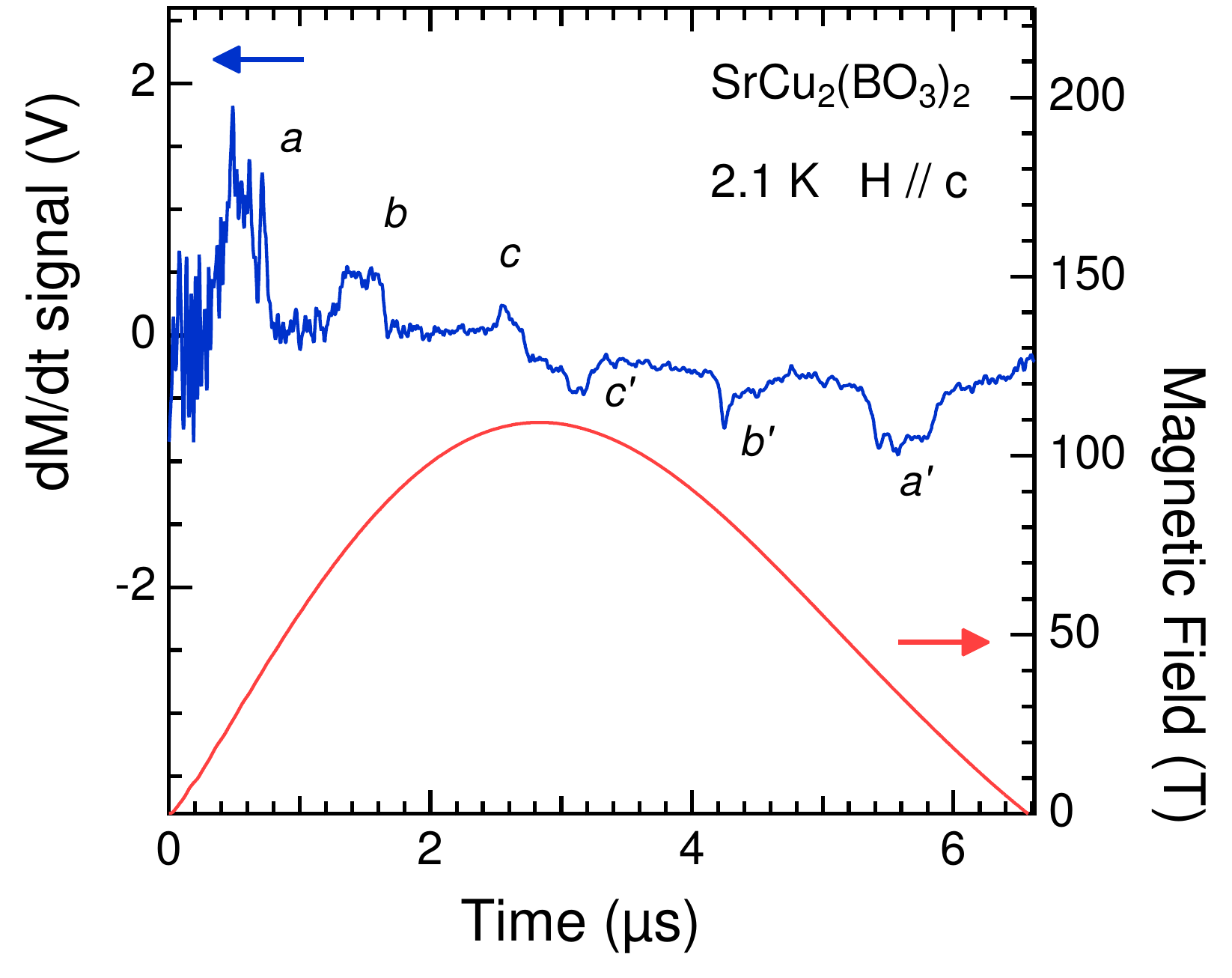}
\caption{\label{fig:dmdt}(Color online) Pickup coil signal proportional to the time derivative of the magnetization ($\text{d}M/\text{d}t$) plotted as a function of time. The magnetic field waveform $H(t)$ is also shown as a solid red line. 
}
\end{figure}

\paragraph{Experimental procedure.---}
A single crystal of SrCu$_2$(BO$_3$)$_2$ was used for the experiment.
Pulsed magnetic fields of up to 118~T were generated by a destructive method; 
the vertical-type single-turn coil technique \cite{TakeMag12JPSJ} was used.
The field was applied parallel to the $c$-axis of the crystal.
The magnetization ($M$) was measured using a pickup coil that consists of two small coils (1 mm diameter, 
1.4 mm length for each).
The two coils have different polarizations and are connected in series. 
The sample is inserted into one of the coils.
An induction voltage proportional to the time derivative of $M$ ($\text{d}M/\text{d}t$) is obtained when the sample 
gets magnetized by a pulsed magnetic field  $H(t)$, where $t$ is the time.
The induction voltage due to $\text{d}H/\text{d}t$ is almost canceled out between the opposite polarization coils. 
The detailed experimental setup for the magnetization measurement using this vertical type single-turn coil 
method has been described elsewhere \cite{TakeMag12JPSJ}. 
A liquid helium bath cryostat with the tail part made of plastic has been used; the sample was immersed in liquid 
helium and a measurement temperature of about 2~K has been reached by reducing the vapor pressure.  

\begin{figure}[b]
\includegraphics[width=1\columnwidth]{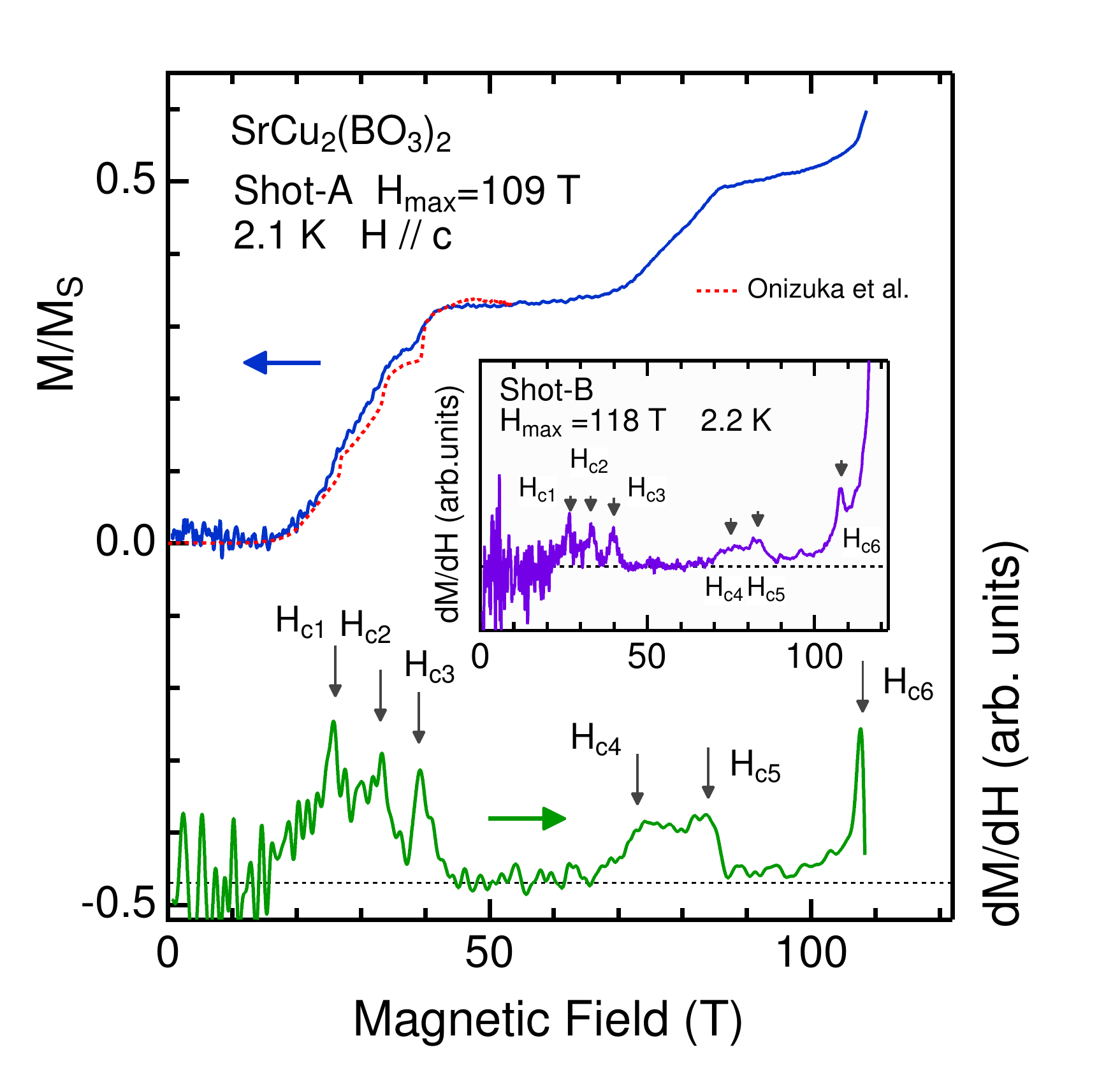}
\caption{\label{fig:mcurve}(Color online)
The magnetization curve at 2.1~K up to 109~T (Shot-A).
Applied fields $H$ are parallel to the $c$-axis of the crystal.
The magnetic field derivative of the magnetization ($\text{d}M/\text{d}H$) curve is displayed 
as a function of magnetic field $H$. The dotted curve is the magnetization curve 
reported previously \cite{Onizuka00}.
$\text{d}M/\text{d}H$ curve of another measurement up to 118~T (Shot-B) is plotted in the inset.
}
\end{figure}

\paragraph{Experimental results.---}
The pickup coil signal proportional to $\text{d}M/\text{d}t$ is shown as a function of time in Fig.~\ref{fig:dmdt} together with 
the magnetic field waveform. 
The obtained maximum field is 109~T and we name this experiment Shot-A in this paper.
Distinct peak structures denoted by labels $a$, $b$, $c$, 
$c^{\prime}$, $b^{\prime}$ and $a^{\prime}$ are present in $\text{d}M/\text{d}t$. 
They correspond to magnetization jumps at the phase boundaries of different spin states. 
Indeed, a stepwise magnetization increase gives rise to a peak in $\text{d}M/\text{d}t$ curve, and the peak is positive (negative) 
for increasing (decreasing) field. 
The one to one correspondence between $a$ and $a^{\prime}$, $b$ and $b^{\prime}$, and $c$ and $c^{\prime}$ indicates 
that stepwise transitions take place at these magnetic fields for both
field-increasing and decreasing processes without significant hysteresis.

The magnetization curve is obtained by a numerical integration of $\text{d}M/\text{d}t$;  
the resulting magnetization $M$ is normalized by the expected saturation magnetization $M_S$. 
The magnetic field derivative of the magnetization $\text{d}M/\text{d}H$ is obtained from the ratio $\text{d}M/\text{d}t \times 1/(\text{d}H/\text{d}t)$.

Figure~\ref{fig:mcurve} shows the magnetization process and the magnetic field dependence of $\text{d}M/\text{d}H$ at 2.1~K (Shot-A). 
We also show for comparison the magnetization $M/M_S$ up to 55~T previously
reported in Ref.~\onlinecite{Onizuka00}, and the agreement is good. 
In the present work, we only analyze the result of the field-increasing process because the magnetic field is 
less homogeneous for the field decreasing process due to the mechanical deformation 
of the single-turn coil and to the background non-linear offset of the signal which disturbs the 
precise measurement \cite{TakeMag12JPSJ}.
The $\text{d}M/\text{d}H$ curve shows clear peaks labeled $H_{cn}$ ($n=1-6$) : $H_{c1,c2, c3}$  are attributed to structure $a$ in Fig.~\ref{fig:dmdt}, 
$H_{c4, c5}$ to structure $b$, and $H_{c6}$ to structure $c$.

We show the $\text{d}M/\text{d}H$ curve obtained from another experiment up to 118~T (Shot-B) in the inset of Fig.~\ref{fig:mcurve}. 
The upward behavior at high fields over 100~T is due to the increase of the background noise:
the noise becomes relatively larger near the top of the magnetic field curve 
because the $\text{d}M/\text{d}t$ signal becomes small when $\text{d}H/\text{d}t$ is small.
Although the background noise makes it difficult to obtain a very precise magnetization curve by a numerical integration for Shot-B,
peaks in $\text{d}M/\text{d}H$ are clearly observed at nearly identical values as for Shot-A.
The obtained peak fields are shown in Table \ref{table:peak}.

\begin{table}[tb!]
\begin{center}
 \caption{Transition magnetic fields $H_{cn}$ obtained from the $\text{d}M/\text{d}H$ peaks for Shot-A and -B. 
 The precision of the magnetic field value is likely to be about $\pm1$~T.}
 \begin{tabular}{ c c c c c c c}
 \hline \hline
    & $H_{c1} (T)$ & $H_{c2} (T)$ & $H_{c3} (T)$ & $H_{c4} (T)$ & $H_{c5} (T)$ & $H_{c6} (T)$  \\ \hline  
Shot-A & 26  & 33   & 39   & 73   & 84   & 108  \\
Shot-B & 27   & 33   & 40   & 75   & 83   & 108    \\
\hline  \hline
 \end{tabular} 
 \label{table:peak}
\end{center}
\end{table}

The $\text{d}M/\text{d}H$ peaks at $H_{c1}$, $H_{c2}$, and $H_{c3}$ are attributed to the magnetization jumps at fields where 
the spin state enters 1/8, 1/4, and 1/3 plateau phases, respectively. Additional
features probably related to extra plateaux \cite{Takigawa12} are also present between $H_{c1}$ and $H_{c2}$, but steady field measurements are more accurate in that field range, and we will not
attempt to discuss them.
While the measurement temperature 2.1~K seems to be too high to observe the 1/8 plateau
\cite{kodama02} the adiabatic cooling owing to 
the first sweep speed of the magnetic field leads to an actual temperature lower than 0.5~K~\cite{Levy08}.
In the magnetization curve, the 1/3 plateau is observed in the field range from 39 to 73~T for Shot-A.
Here note that we calibrate the absolute value of $M/M_S$ using the magnetization at the 1/3 plateau phase.
The field region for the 1/3 plateau is in good agreement with the previous reports
\cite{Onizuka00,Sebastian08}.

After the 1/3 plateau, there is a change of slope around 74~T. Above that critical field $H_{c4}$, there is an almost
smooth increase of the magnetization, followed by the appearance of the 1/2 plateau at around $H_{c5} \simeq 84~T$. 
Note however that a trapezoid or broad flap-top peak is expected if the slope increase was monotonous and had no anomaly. 
Since a peak structure is clearly observed both in up and down sweeps between the 1/3 and 1/2 plateau (see in particular
feature $b'$ in down sweep), some kind of transition probably takes place between the 1/3 and 1/2 plateaux. 

The 1/2 plateau starts at 84~T and continues up to 108~T.
The starting magnetic field seems to be slightly higher compared to the previously reported value around 82~T detected by  
magnetostriction \cite{Jaime12}.
This might be partly due to the different ways of detection (magnetostriction versus magnetization), and also to the experimental 
uncertainty in the present work (the error of the 
absolute value of the magnetic field is within 3\%).
The magnetic field absolute value of the single-turn coil method contains a few percent experimental
error owing to the technical limit of the precision \cite{TakeMag12JPSJ}.  
However, even if there is an error bar on the absolute value of the magnetic field, the relative change in the field value has a smaller error bar. 
Hence it is safe to conclude that the plateau length of the 1/2 plateau $\Delta H \simeq 24$ ~T is considerably shorter than that of 1/3 plateau $\Delta H \simeq 34$~T.
At higher fields, considering the appearance of a sharp peak $H_{c6}$, a first-order magnetic phase transition is expected to occur 
after the 1/2 plateau at a field of 108~T.


\begin{figure}[b]
\includegraphics[width=1\columnwidth]{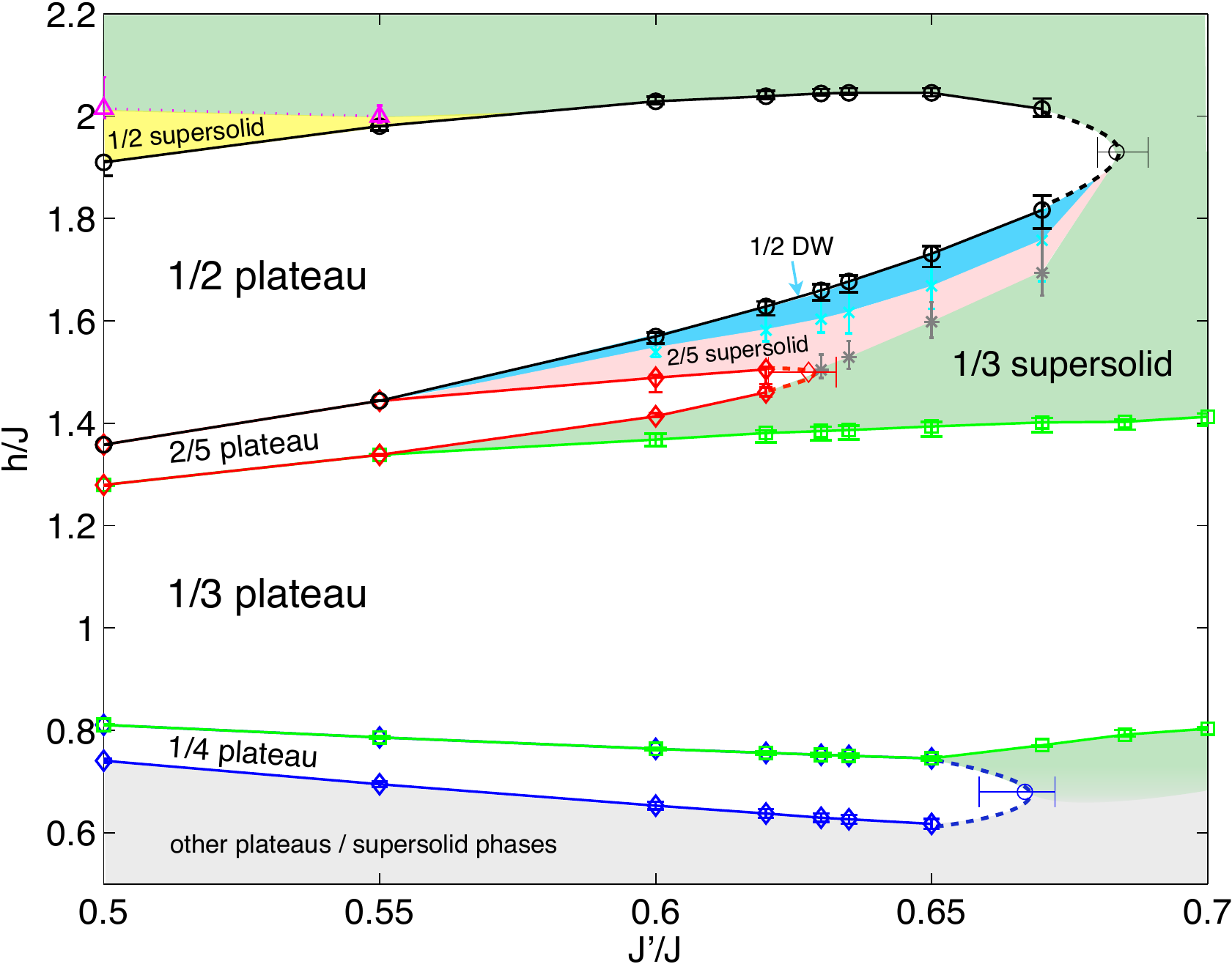}
\caption{\label{fig:ipeps}(Color online) Phase diagram of the Shastry-Sutherland
model in a magnetic field obtained with iPEPS.
}
\end{figure}

\paragraph{Theory.---}
A good starting point to describe the magnetization process of SrCu$_2$(BO$_3$)$_2$ is 
provided
by the spin-1/2 Heisenberg model on the Shastry-Sutherland lattice defined by:
\begin{equation}
H=J'\sum_{<i,j>}\bm S_{i}\cdot \bm S_{j}+J\sum_{\ll i,j\gg}\bm
S_{i}\cdot \bm S_{j}-h\sum_{i}S_i^z
\end{equation}
where the $\ll i,j\gg$ bonds with coupling $J$ build an array of orthogonal dimers while
the $<i,j>$ bonds with coupling $J'$ denote inter-dimer couplings. While a lot of effort has been 
devoted in the past to the magnetization curve up to $1/3$ \cite{dorier08,abendschein08,nemec12}, in the range where a sequence of plateaux 
has been reported, comparatively little attention has been paid so far to the magnetization curve
above $1/3$. Shortly after the discovery of plateaux in SrCu$_2$(BO$_3$)$_2$, Momoi and 
Totsuka \cite{MT00PRB} have predicted the presence of $1/3$ and $1/2$ plateaux separated by supersolid
phases. This prediction has been left unchallenged until the recent investigation of
magnetostriction in very high field \cite{Jaime12}. These measurements have revealed the presence of
an anomaly above the $1/3$ plateau that has been interpreted as a $2/5$ plateau, 
an interpretation backed by a DMRG (density matrix renormalization group) calculation 
at $J'/J=0.62$. However, a recent tensor-network calculation based on MERA (multi-scale
entanglement renormalization ansatz) has just confirmed the presence of $1/3$ and $1/2$
plateaux without any evidence of a $2/5$ plateau \cite{lou12}.

In view of the importance of this issue for the interpretation of the present results, we have 
decided to reinvestigate the high-field magnetization process of the Shastry-Sutherland model 
with a variety of state-of-the-art
numerical approaches: exact diagonalizations of finite-size clusters up to 40 spins, DMRG
on clusters of size up to $12 \times 10$ spins, high-order series expansions,
and iPEPS -- a tensor network method for two-dimensional systems in the thermodynamic limit. The 
various methods yield a rather consistent picture (see supplemental material for a detailed 
comparison). The most complete phase diagram, shown in Fig.~\ref{fig:ipeps}, has been obtained with iPEPS. Above the $1/3$ plateau, it consists of
two additional plateaux at $2/5$ and $1/2$, three supersolid phases with the symmetries of the
$1/3$, $2/5$ and $1/2$ plateaux, and a phase with domain walls separating regions of $1/2$
plateau structures. Note that we confirm the presence of a 2/5 plateau for $J'/J=0.62$, in agreement
with the DMRG results of Ref.~\onlinecite{Jaime12}.

\begin{figure}[b]
\includegraphics[width=1\columnwidth]{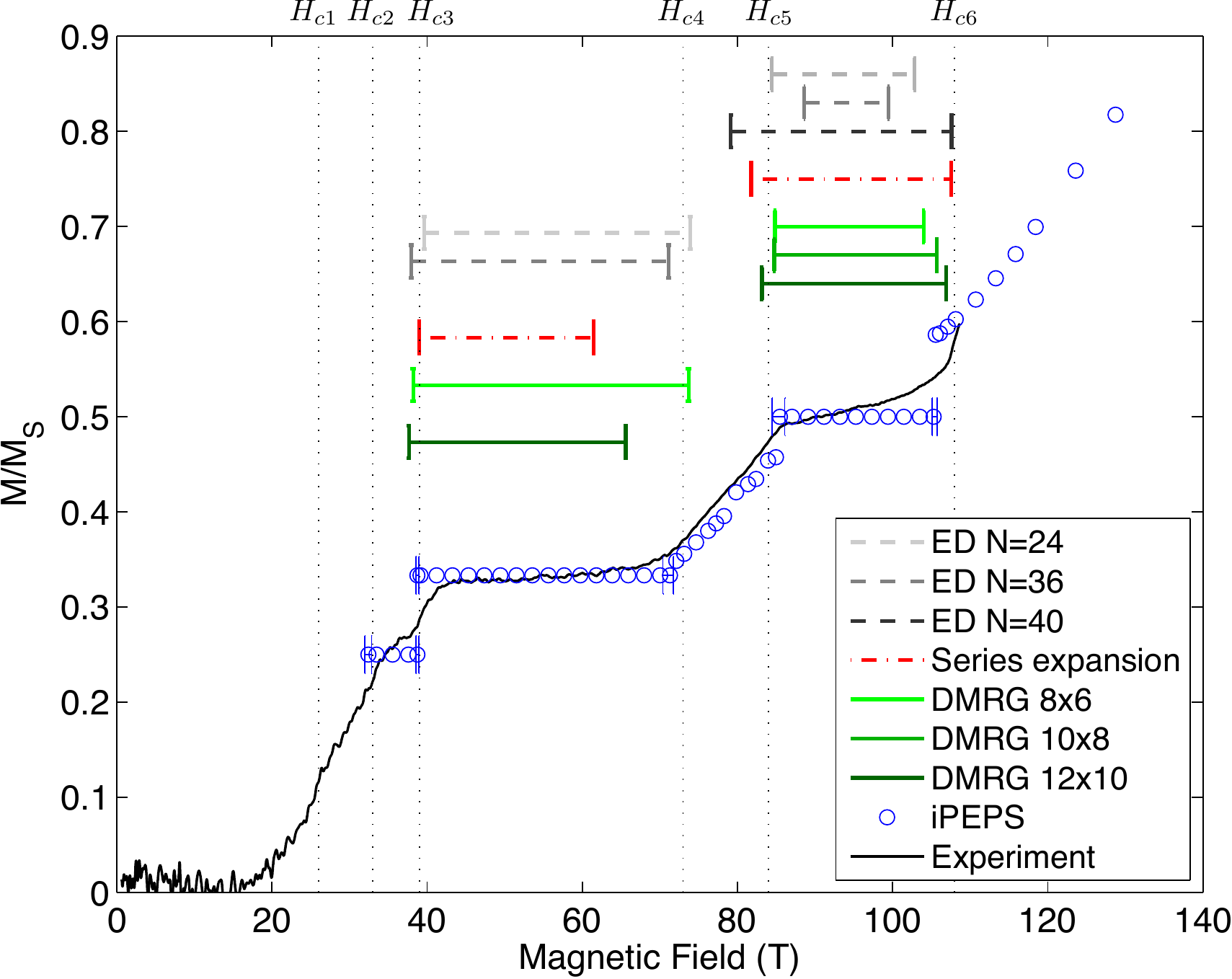}
\caption{\label{fig:expt_theory}(Color online) Comparison between the experimental magnetization curve and the iPEPS simulation results for $J'/J=0.63$. The extent of the 1/3 and 1/2 plateaux predicted by the other methods is shown on top of the plateaux. 
}
\end{figure}

For our present purpose, the most important messages of this phase diagram are: i) The 1/2 plateau
does not extend beyond a critical value of the order of \mbox{$J'/J\simeq0.685$}, in qualitative agreement
with Momoi and Totsuka \cite{MT00PRB}; ii)~The $2/5$ plateau does not
extend beyond $J'/J\simeq 0.625$. Since the present experimental data do not reveal any evidence of a $2/5$ 
plateau but show a rather broad $1/2$ plateau, $J'/J$ can neither be too large nor too small, and 
a  comparison of the critical fields of the $1/2$ and $1/3$ plateaux with the experimental ones point to a ratio $J'/J\simeq0.63$.

A detailed comparison of the experimental magnetization curve with the theoretical predictions
of the various methods at $J'/J\simeq0.63$ above the 1/4 plateau is shown in 
Fig.~\ref{fig:expt_theory}. First of all, the critical fields $H_{c3}$ to  $H_{c6}$
are accurately reproduced by iPEPS. The predictions of the other methods are scattered around
the iPEPS values, but altogether they support the main features of the iPEPS results 
(for a detailed comparison
as a function of $J'/J$, see supplemental material). 
Secondly, the magnetization jumps at $H_{c3}$ and $H_{c6}$, which point to first-order
transitions, are well accounted for by the theoretical results: at $H_{c3}$, there is a first-order 
transition between the 1/4 and 1/3 plateau, while at $H_{c6}$, there is one between the 1/2 plateau
and the 1/3 supersolid. The smoother transitions at $H_{c4}$ and $H_{c5}$  also
correspond to much weaker anomalies in the theoretical results. For the upper boundary of the 1/3 plateau, 
series expansions point to a gap closing when increasing $H$, hence to a second
order phase transition, around $65$~T, 
significantly below $H_{c4}$. This is not incompatible with the broad onset of magnetization
around $H_{c4}$, with a slope that takes off around $65$~T in shot-A and $70$~T in shot-B.
Below the lower boundary of the 1/2 plateau at $H_{c5}$, iPEPS predicts a series of first order 
phase transitions from a 1/3 supersolid to a 2/5 supersolid, then to a phase with domain walls, and then
finally to the 1/2 plateau. In the magnetization curve, these transitions translate into small jumps.
This is presumably related to the peak observed  in both shots around $80$~T,
i.e., between the 1/3 and 
1/2 plateaux, consistent with the prediction
that the intermediate field range between these plateaux is not a single phase.

Finally, let us comment on the experimental slope of the 1/2 plateau between $H_{c5}$ and $H_{c6}$,
which is anomalously large as compared, e.g., to that of the 1/3 plateau. This slope is definitely too large
to be due to Dzyaloshinskii-Moriya interactions, but it might be simply explained as a temperature
effect. Indeed, the difference in energy per spin between the 1/2 plateau and the competing 1/3 supersolid state obtained with iPEPS is very small (
$<0.004 J$), whereas the competing phases are definitely  higher in the middle of the 1/3 plateau.

\paragraph{Conclusion.---}
To summarize, we have performed ultra-high field measurements of the magnetization of SrCu$_2$(BO$_3$)$_2$,
revealing for the first time the extent of the 1/2 plateau.
The length of the 1/2 plateau has been found to be around 70\%\ of that of the 1/3 plateau.
We have not found any indication of the 2/5 plateau
that was previously suggested on the basis of magnetostriction measurements. As revealed by large-scale numerical simulations,
these results are consistent
with the Shastry-Sutherland model provided the ratio of inter to intra-dimer coupling is neither too small, in agreement with
recent results on Zn doped samples~\cite{Yoshida}, nor too
large,  the best agreement being reached for a ratio of about $0.63$.
These numerical simulations further predict that the magnetization between the 1/3 and 1/2 plateau and above the 1/2 plateau
is not uniform, but that the system is always in a phase that breaks the translational symmetry, either to form
a supersolid, or because of the spontaneous appearance of domain walls in the 1/2 plateau phase. 
It would be very interesting to test this prediction with measurements that can detect a change of lattice symmetry
such as X-rays or neutrons, or with a local probe such as NMR. Given the field range of interest, this is however a huge
experimental challenge. 

\paragraph{Acknowledgement.---}
Y.~H.~M.\ thanks M. Takigawa for fruitful discussions. A.~H.\ and P.~C.\ acknowledge support through FOR1807 (DFG / SNSF).
We acknowledge allocation of CPU time at the HLRN Hannover. The iPEPS simulations have been performed on the Brutus cluster at ETH Zurich.

\bibliography{SCBO}

\end{document}


\title{Magnetization of SrCu$_2$(BO$_3$)$_2$ in ultrahigh magnetic fields up to 118~T: \textit{supplemental material} }

\author{Y.~H.~Matsuda}
\affiliation{Institute for Solid State Physics, University of Tokyo, Kashiwa, Chiba 277-8581, Japan}

\author{N.~Abe}
\affiliation{Institute for Solid State Physics, University of Tokyo, Kashiwa, Chiba 277-8581, Japan}

\author{S.~Takeyama}
\affiliation{Institute for Solid State Physics, University of Tokyo, Kashiwa, Chiba 277-8581, Japan}

\author{H.~Kageyama}
\affiliation{Graduate School of Engineering, Kyoto University, Nishikyouku, Kyoto 615-8510, Japan}

\author{P.~Corboz}
\affiliation{Theoretische Physik, ETH Z\"urich, CH-8093 Z\"urich, Switzerland}

\author{A.~Honecker}
\affiliation{Institut f\"ur Theoretische Physik, Georg-August-Universit\"at G\"ottingen,
     D-37077 G\"ottingen, Germany}
\affiliation{Fakult\"at f\"ur Mathematik und Informatik, Georg-August-Universit\"at G\"ottingen,
     D-37073 G\"ottingen, Germany}

\author{S.~R.~Manmana}
\affiliation{Institut f\"ur Theoretische Physik, Georg-August-Universit\"at G\"ottingen,
     D-37077 G\"ottingen, Germany}

\author{G.~R.~Foltin}
\affiliation{Lehrstuhl f\"{u}r Theoretische Physik I, Otto-Hahn-Stra\ss e 4, TU Dortmund, 44221 Dortmund, Germany}

\author{K.~P.~Schmidt}
\affiliation{Lehrstuhl f\"{u}r Theoretische Physik I, Otto-Hahn-Stra\ss e 4, TU Dortmund, 44221 Dortmund, Germany}

\author{F.~Mila}
\affiliation{Institute of Theoretical Physics, Ecole Polytechnique F\'ed\'erale de Lausanne (EPFL), 1015 Lausanne, Switzerland}

\date{August 19, 2013}

\maketitle

This supplemental material is organized as follows: In Secs.~\ref{sec1}-\ref{sec3} details on the experimental  techniques for the generation of ultrahigh magnetic fields over 100~T and the 
magnetization measurements are given. In Sec.~\ref{sec:methods} we provide an overview of the numerical methods used in our study of the Shastry-Sutherland model. In Sec.~\ref{sec:phases} we present the spin structures of the phases mentioned in the main text. Finally, in Sec.~\ref{sec:comp} we compare the different numerical results for the extent of the 1/3 and 1/2 plateau phases, and discuss magnetization curves obtained for different values of $J'/J$ in comparison with the experimental data.

\section{Single-turn coil technique}
\label{sec1}
The generation of a strong magnetic field exceeding 100~T is technically very difficult because of the 
huge Maxwell force. A great deal of effort has been done to extend the field range; a record of a 
magnetic field of 100.75~T was recently obtained in nondestructive manner
\cite{Jaime12}. 
However, it is widely recognized that a magnetic field far above 100~T is only generated in a destructive
manner, i.e., by destroying the magnet. The electromagnetic flux compression (EMFC) method allows 
us to generate high fields over 700~T \cite{takeyama11,miyata11}.
However, since everything inside the magnet including the sample are completely destroyed in 
the EMFC experiment, substantial effort is required to repeat the experiment.

\begin{figure}[b!]
\includegraphics[width=8cm]{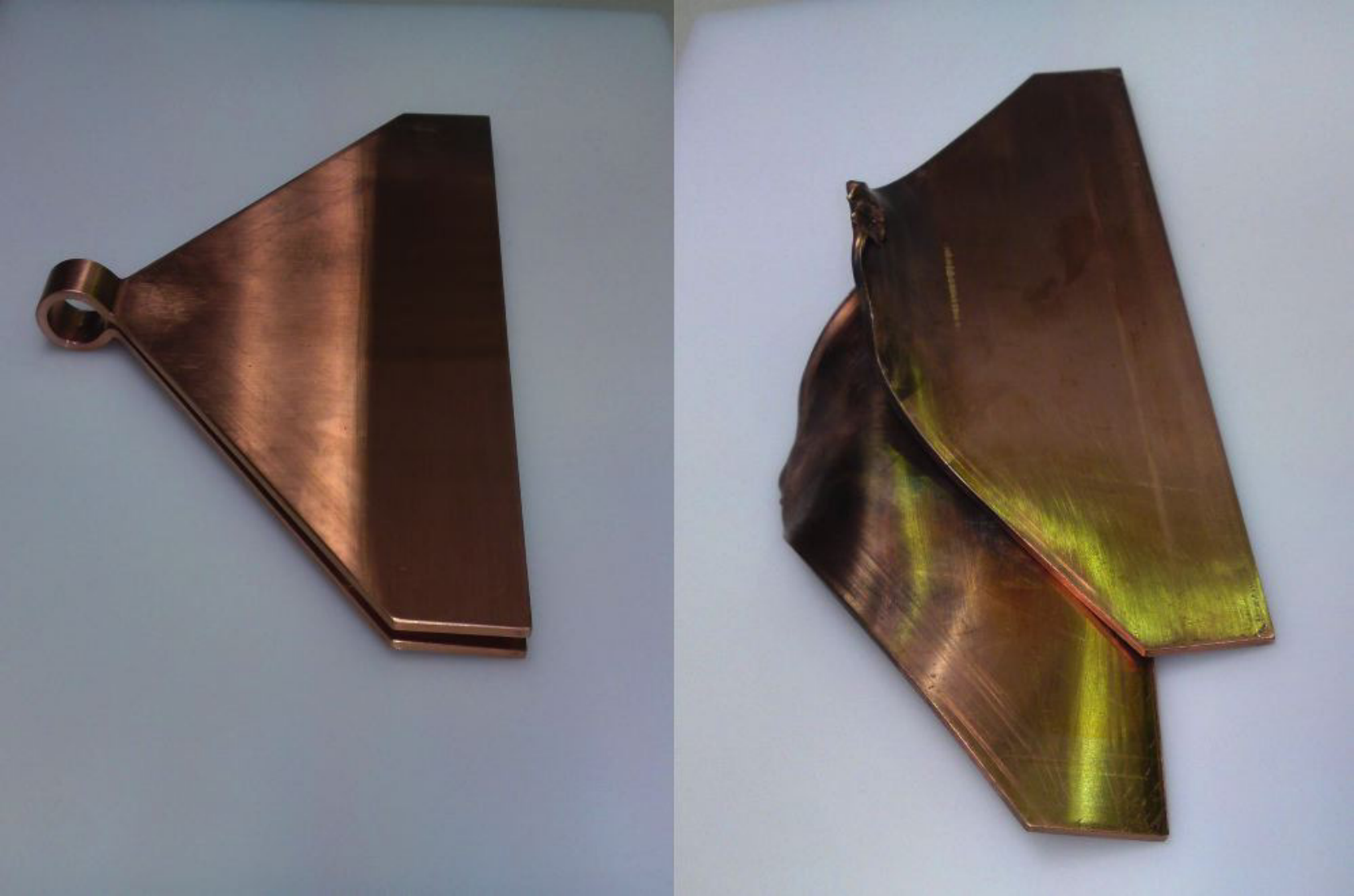}
\caption{\label{fig:coil} A single-turn coil before the experiment (left) and after the experiment (right). The inner 
diameter of the coil in this picture is 14 mm.}
\end{figure}

The single-turn coil (STC) technique is useful for high magnetic field experiments up to 200~T.
Although the magnet (single-turned coil) is broken in the field generation process, the sample and the cryostat 
inside the coil survives with almost no damage \cite{nakao}.  The coil for the STC technique is lightweight and rather small as shown 
in Fig.~\ref{fig:coil}. Hence the experiment is much easier than EMFC.

\begin{figure}[t!]
\includegraphics[width=7cm]{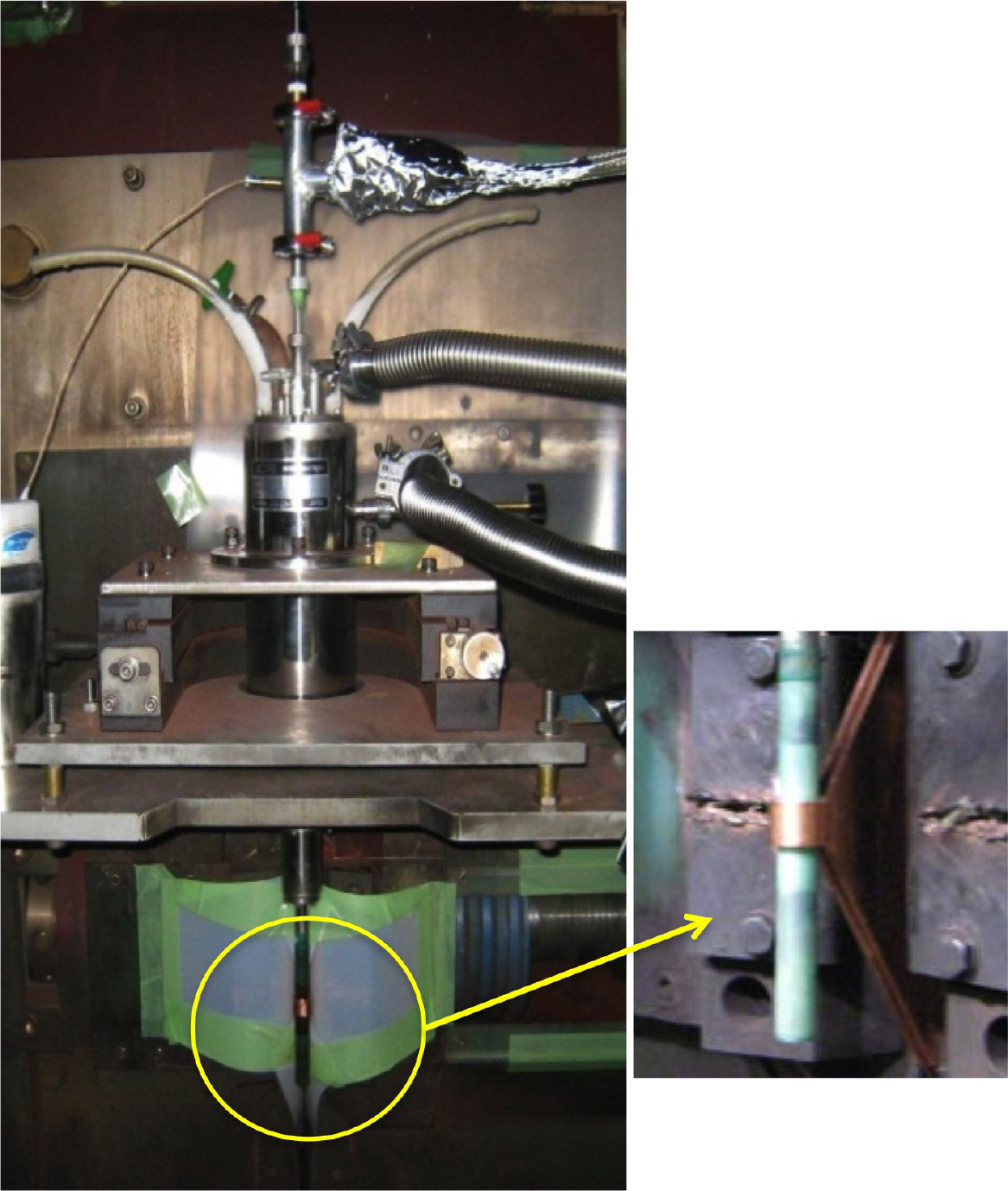}
\caption{\label{fig:vstc} The vertical type single-turn coil apparatus; a helium bath cryostat is installed. 
A single-turn coil with 14 mm diameter is inserted.
}
\end{figure}

The vertical type STC in the Institute for Solid State Physics, University of
Tokyo \cite{miura03},  was utilized in the present study.
The coil is vertically set to the electrode so that a liquid-helium bath
cryostat is inserted into the coil bore.
The photo of the set-up of the coil and the cryostat is shown in Fig.~\ref{fig:vstc}. 
A capacitor bank is used as the power source; the full electrical capacitance is 263.5~$\mu$F and the
maximum charging voltage is 40~kV.
The typical waveform of the generated magnetic field using a single-turn coil with 14~mm diameter is shown in Fig.~1 in the main text. 

\section{A helium bath cryostat specially designed for the vertical STC}
\label{sec2}
\begin{figure}[b!]
\includegraphics[width=7cm]{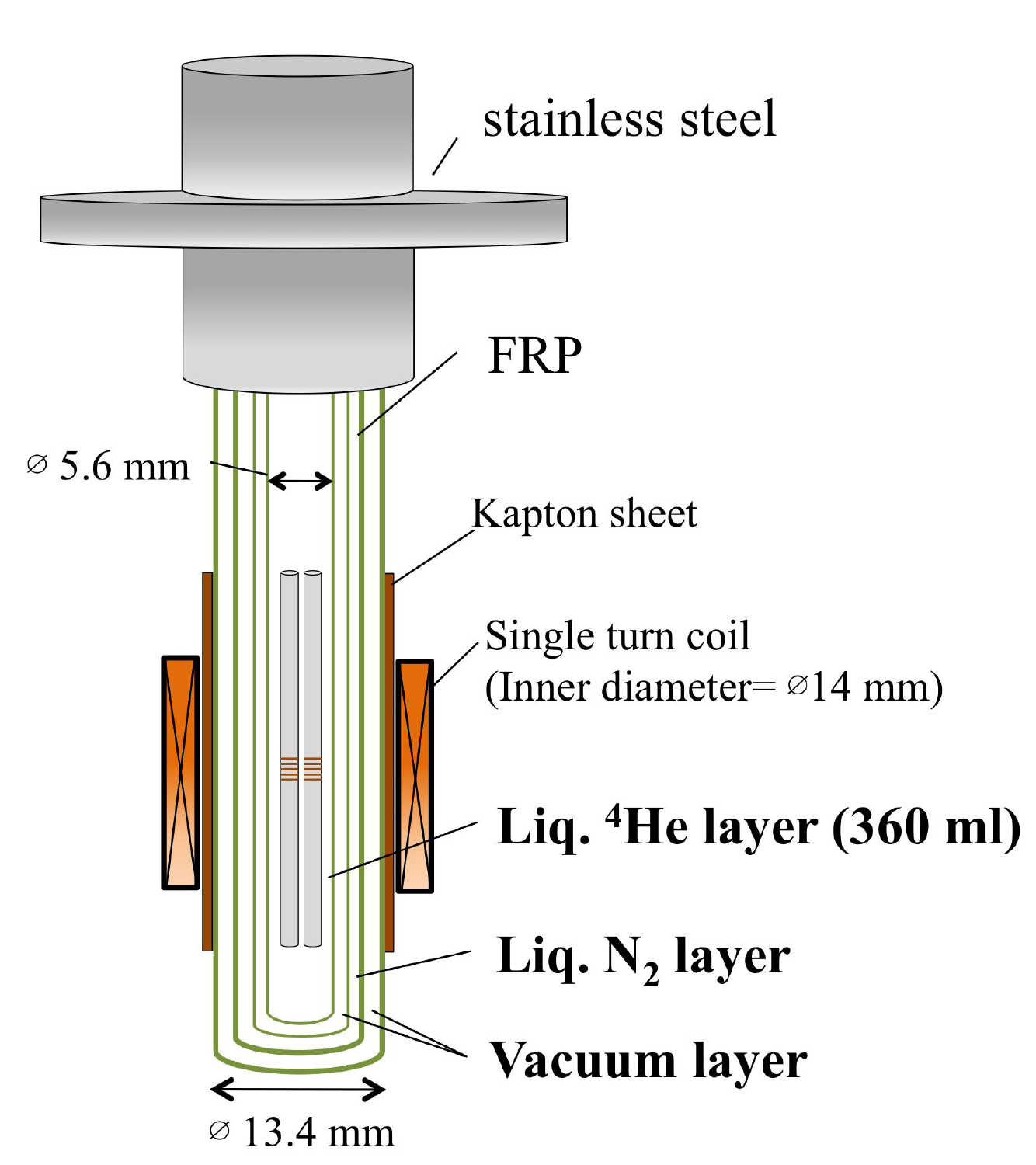}
\caption{\label{fig:cryo} Schematic diagram of  the cryostat specially designed for the vertical STC.
}
\end{figure}

The duration time of the magnetic field generated by the STC is 6 -- 9 $\mu$s. Such a high speed pulsed magnetic field with an intense peak value 
larger than 100~T gives rise to a large induction current in metals located near by the coil.
For instance, a metal tube inside the coil would be strongly deformed and might cause an implosion due to the 
strong electromagnetic force between the induction current and the magnetic field.
Therefore a helium bath cryostat with the tail section made of a
fiber-reinforced glass epoxy (so-called FRP or G10) was specially designed
\cite{TakeMag12JPSJ}.
The schematic diagram of the cryostat is shown in Fig.~\ref{fig:cryo}. 
The sample is immersed in liquid helium.
A low temperature down to 2~K is reached by evacuation of helium vapor. 

\section{Magnetization measurement using the STC}
\label{sec3}
\begin{figure}[t!]
\includegraphics[width=7cm]{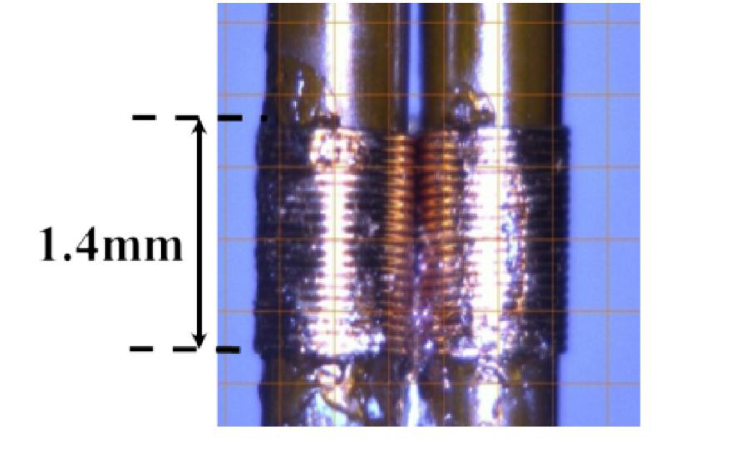}
\caption{\label{fig:pickup} Photograph of the pickup coil for the magnetization measurement .
}
\end{figure}

The magnetization measurement was performed using a pair of  pickup coils as shown in Fig.~\ref{fig:pickup}.  
It is wound 20 turns around a polyimide tube (with an outer diameter of 1.12 mm) by a copper wire (with a diameter of 60 $\mu$m) for each coil.
The coils are series-connected and have opposite polarization so that the induction voltage by an applied magnetic field is canceled.
When a sample is inserted into the right pickup coil (R-coil) while the left coil (L-coil) remains empty, the signal induced in the R-coil ($V_{\textrm{R}}$) 
and that in the L-coil ($V_{\textrm{L}}$) are described as 
\begin{eqnarray}
\label{eq:induction1}
V_\textrm{R}=S_{\textrm{eff}}^A \mu_0 \frac{dH}{dt} + S_{\textrm{eff}}^A \frac{dM}{dt}  \\  \nonumber
V_\textrm{L}=-S_{\textrm{eff}}^B \mu_0 \frac{dH}{dt} .
\end{eqnarray}
Here, $S_{\textrm{eff}}^A$ and $S_{\textrm{eff}}^B$ are the effective area of the pickup coils (including the number of turns), respectively, 
and $\mu_0$ is the magnetic permeability of the vacuum.  $H$ denotes the applied magnetic field.
The obtained signal in the experiment is 
\begin{eqnarray}
\label{eq:induction2}
V_1=V_R+V_L=(S_{\textrm{eff}}^A-S_{\textrm{eff}}^B ) (\mu_0 \frac{dH}{dt})+ S_{\textrm{eff}}^A \frac{dM}{dt}. 
\end{eqnarray}
A great deal of effort is done to make the coils such that $S_{\textrm{eff}}^A \sim S_{\textrm{eff}}^B$ and the condition that 
$(S_{\textrm{eff}}^A -S_{\textrm{eff}}^B)/ S_{\textrm{eff}}^A  \sim 10^{-4}$ is required for precise measurements.
This is because the induction voltage for each coil ($S_{\textrm{eff}}^{A (B)}\mu_0 \frac{dH}{dt}$) can become as high as 1000~V. 

The first term of eq.~(\ref{eq:induction2}) is the background noise owing to the imperfect compensation between the R- and L-coils.
This is further canceled by repeating the measurement with the condition that the sample position is exchanged from the R-coil to L-coil.
The signal obtained in the second measurement is 
\begin{eqnarray}
\label{eq:induction3}
V_2=V_R+V_L=(S_{\textrm{eff}}^A-S_{\textrm{eff}}^B ) (\mu_0 \frac{dH}{dt})- S_{\textrm{eff}}^B \frac{dM}{dt}.
\end{eqnarray}
Finally, the signal that is proportional to the magnetization is obtained as follows,
\begin{eqnarray}
\label{eq:induction4}
V=V_1-V_2=(S_{\textrm{eff}}^A+S_{\textrm{eff}}^B )  \frac{dM}{dt} \sim 2 S_{\textrm{eff}}^A \frac{dM}{dt} .
\end{eqnarray}
The $dM/dt$ signal plotted in Fig.~1 in the main text was deduced from eq.~(\ref{eq:induction4}).


\section{Numerical Methods}
\label{sec:methods}
\subsection{Exact diagonalization}

Exact diagonalization (ED) using the Lanczos method is a versatile tool for studying low-dimensional quantum models
(see, e.g., Ref.~\onlinecite{Lauchli11}). ED
has also been widely applied to the $S=1/2$
Shastry-Sutherland model. Nevertheless, to the best of our knowledge,
there are only two publications where exact diagonalization results
on Shastry-Sutherland lattices with more than 32 spins have
been reported \cite{2Dqmag,abendschein08}:
Ref.~\onlinecite{2Dqmag} has presented a
magnetization curve at $J'/J = 0.6$  for $N=32$ and $36$ and
Ref.~\onlinecite{abendschein08} has studied the phase
diagram in a magnetic field using exact diagonalization for $N=32$ and $36$,
but only for $J'/J \le 0.5$.

\begin{figure}[t!]
\begin{center}
\includegraphics[width=0.30\columnwidth]{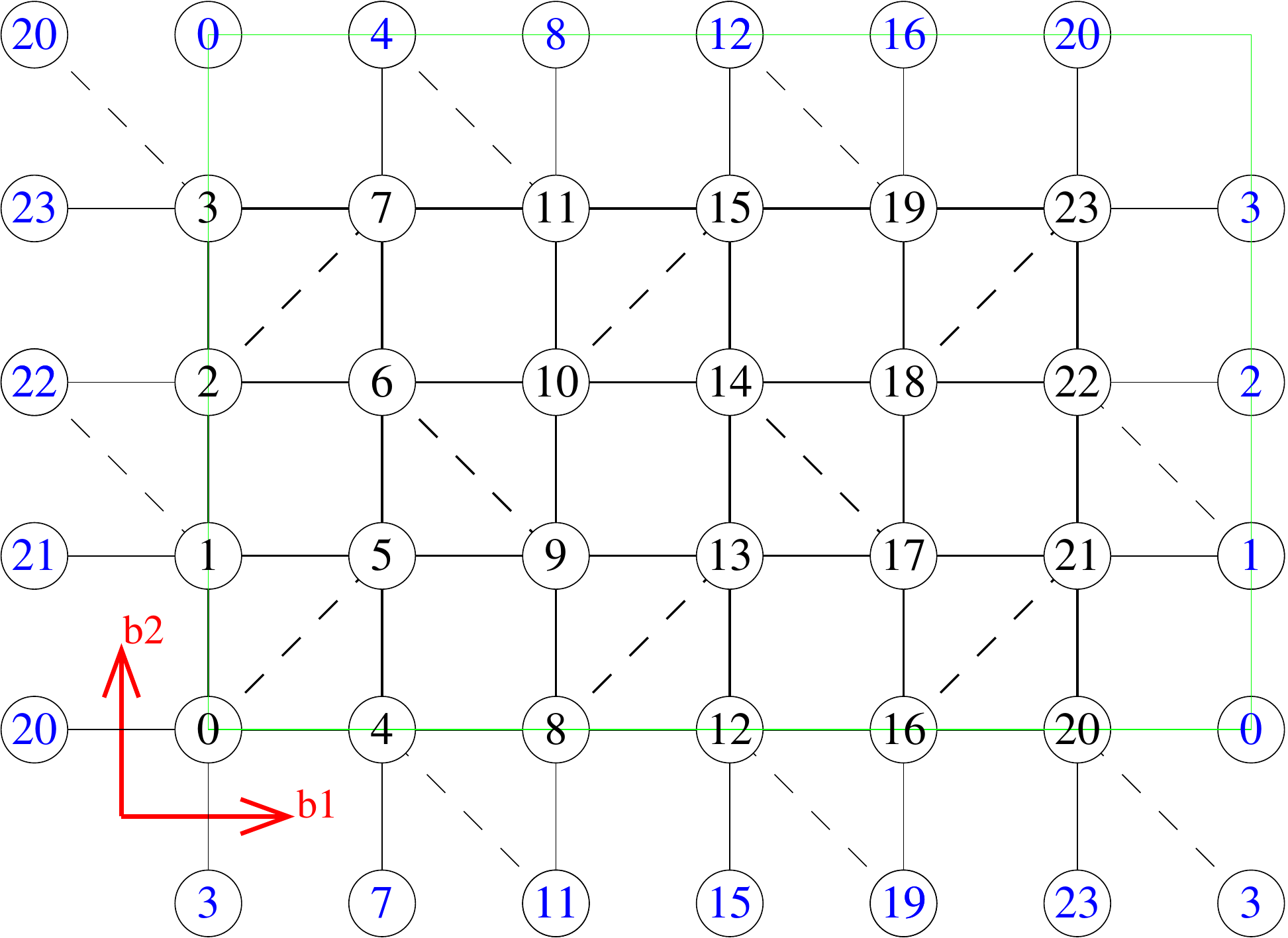}\hfill%
\includegraphics[width=0.31\columnwidth]{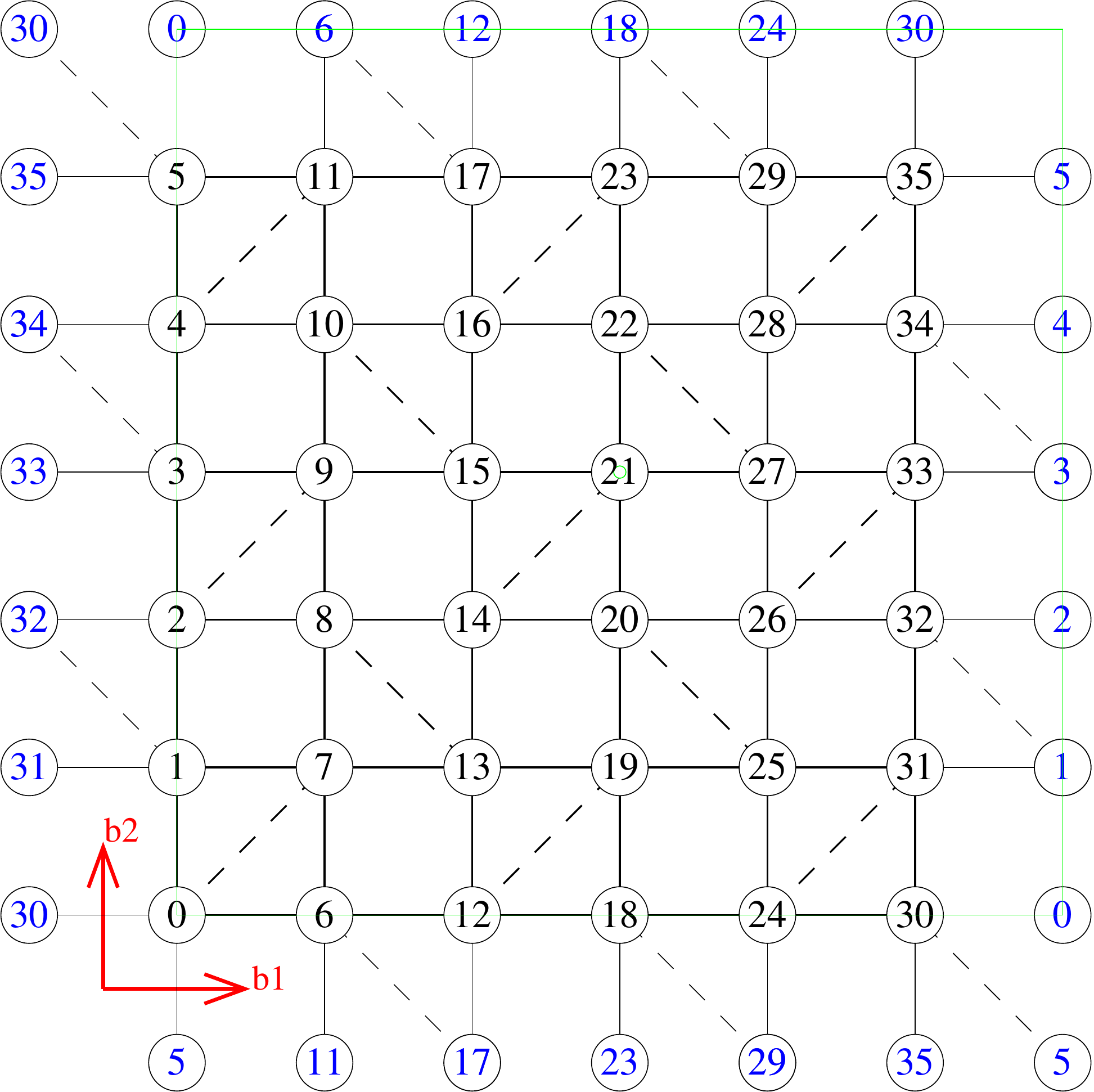}\hfill%
\includegraphics[width=0.32\columnwidth]{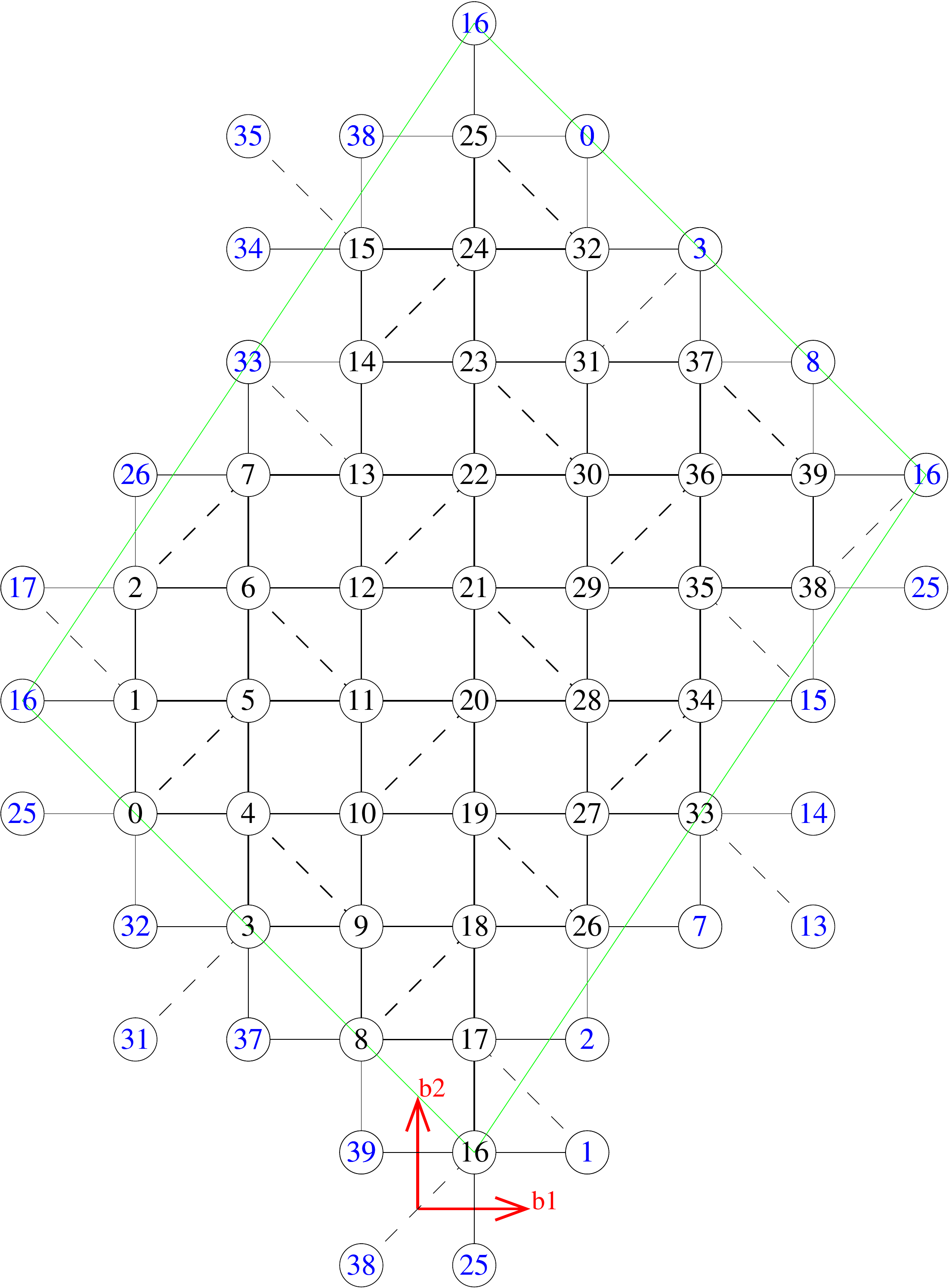}
\end{center}
\caption{From left to right: The $N=24$, 36, and 40 sites Shastry-Sutherland lattices used for
ED in the present work. 
The $6\times6$ $=36$ site lattice is also used for the DMRG simulations. The larger
systems used with DMRG are obtained starting from this lattice and adding an even number of columns and rows, respectively. 
\label{fig:EDlattice}}
\end{figure}

The present computations have been performed with SPINPACK
\cite{Spinpack}. We have employed periodic boundary conditions in order
to minimize finite-size effects and to permit using translational symmetries.
In addition, we have used point group symmetries and
conservation of total $S^z$. Still, Hilbert space dimensions remain
comparably large due to the big unit cell of the Shastry-Sutherland
lattice which contains 4 spins. Therefore, we are rectricted to lattices
with $N \lesssim 40$ spins even if we use MPI parallelization on up to 1536
cores.

Due to the limited system sizes, attention must be paid to finite-size effects
when interpreting ED data.
Firstly, one should keep in mind that the $T=0$ magnetization curve for a finite
lattice consists of at most $N/2$ steps (for $M \ge 0$) and only
magnetizations $M/M_S = 2\,n/N$ with $n$ integer and $\left|n\right| \le N/2$
are realized. Additional finite-size effects will arise if the structure of the
ground state is incommensurate with the lattice under consideration. The
lattices for which we present data are shown in Fig.~\ref{fig:EDlattice}.

\subsection{Density matrix renormalization group}

The density matrix renormalization group method (DMRG) and related matrix
product state (MPS) approaches are standard tools for treating
(quasi-)one-dimensional systems, in particular also spin systems in magnetic
fields \cite{schollwoeck2005,schollwoeck2011}, and recently it has been applied
successfully to two-dimensional systems \cite{stoudenmire2D,white_science,depenbrock2012,balents2012}.
Here, we attempt to characterize the phase diagram of the 2D Shastry-Sutherland model at all values of the magnetization, which is a far more challenging task than analyzing the ground state at $M=0$ only.
Since we are tackling the problem in a combination of methods, using the DMRG we focus solely on the magnetization curve at a few values of $J'/J$ in order to support the results obtained by iPEPS and to the degree possible extend the analysis of finite clusters performed by ED to larger system sizes.  
This is achieved by computing the ground state energy for systems with periodic boundary conditions (PBC) in both spatial directions and cluster sizes of $6\times6$, $8\times6$, $8\times8$, $10\times8$, $10\times10$, and $12\times10$ spins.
Due to the PBC, boundary effects on the energy are avoided. 
Typically, we obtain the energies per site at all values of the magnetization with an accuracy of the order of $5\cdot10^{-3}$ or better (in typical ground state calculations in one-dimensional systems, an accuracy in the energy per site of the order of $10^{-9}$ can be achieved).
However, it is difficult to guarantee that the DMRG does not get stuck in excited states, as comparison with ED data for clusters with $6\times6$ sites at larger values of $J'/J$ than the ones discussed here has shown. 
In principle, this uncertainty can lead to artifacts in the magnetization curve which, however, should not appear in a systematic way throughout the data obtained for different system sizes.
Therefore, if we identify a signature for a plateau in different system sizes, we interpret the finding in that way that the DMRG in these cases converged within the aforementioned accuracy to the correct state. 

Usually, converging the energy at this low accuracy leads to wave functions which can be qualitatively wrong, so that local observables and correlation functions can show the wrong behavior.
However, due to the U(1) symmetry of the system, for computing the magnetization curves we only need the energies of the ground states in all sectors of $S^z_{\rm total}$, which is only a single number per run and also the most accurate observable obtained by the DMRG, since it is a variational method. Thus, despite the difficulties to reach better convergence, we can apply PBC in both spatial dimensions, which is the most challenging scenario for the DMRG. Together with the estimate of the error bars presented in this section, this allows us to obtain the magnetization curve with a good precision, so that we can compare to results obtained by the other approaches. 

\begin{figure*}[t!]
\includegraphics[width=0.49\textwidth]{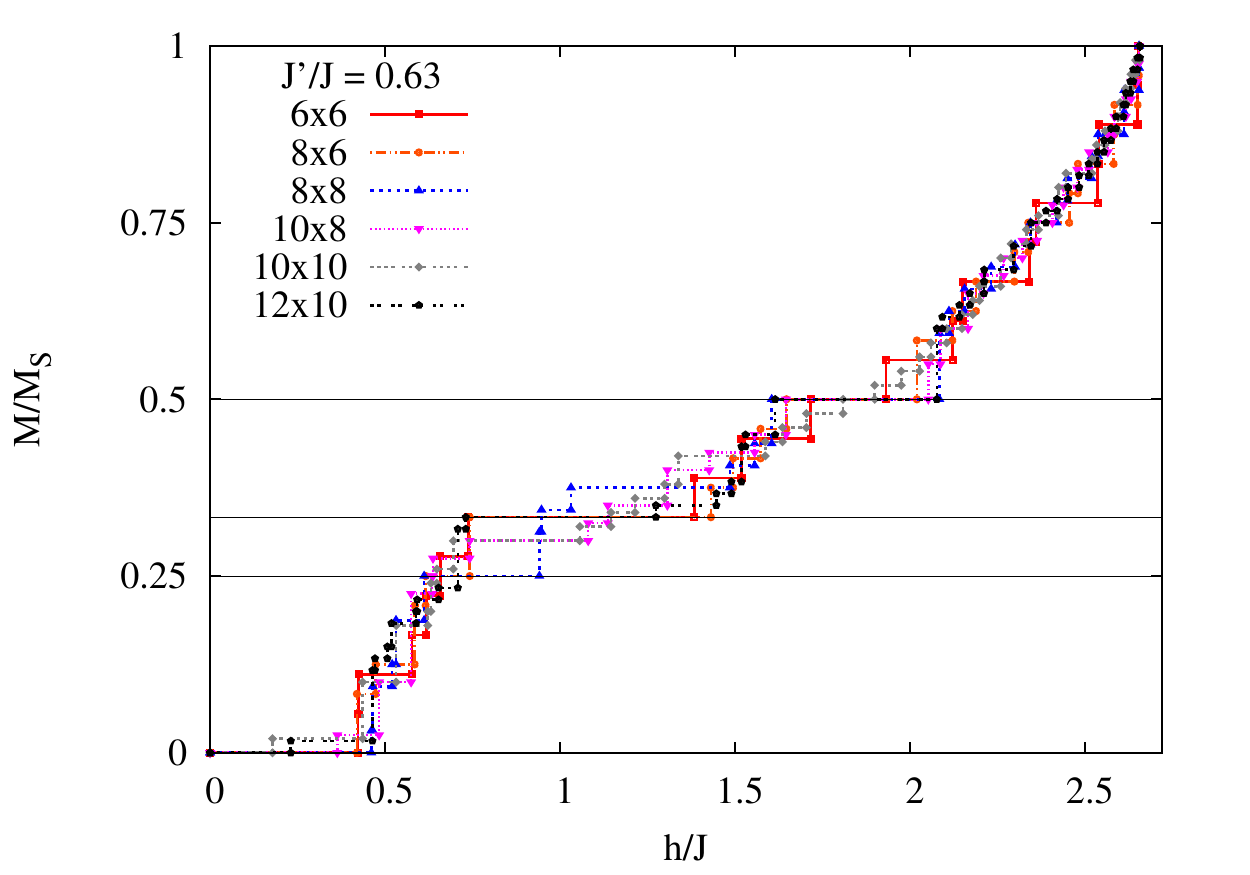}
\includegraphics[width=0.463\textwidth]{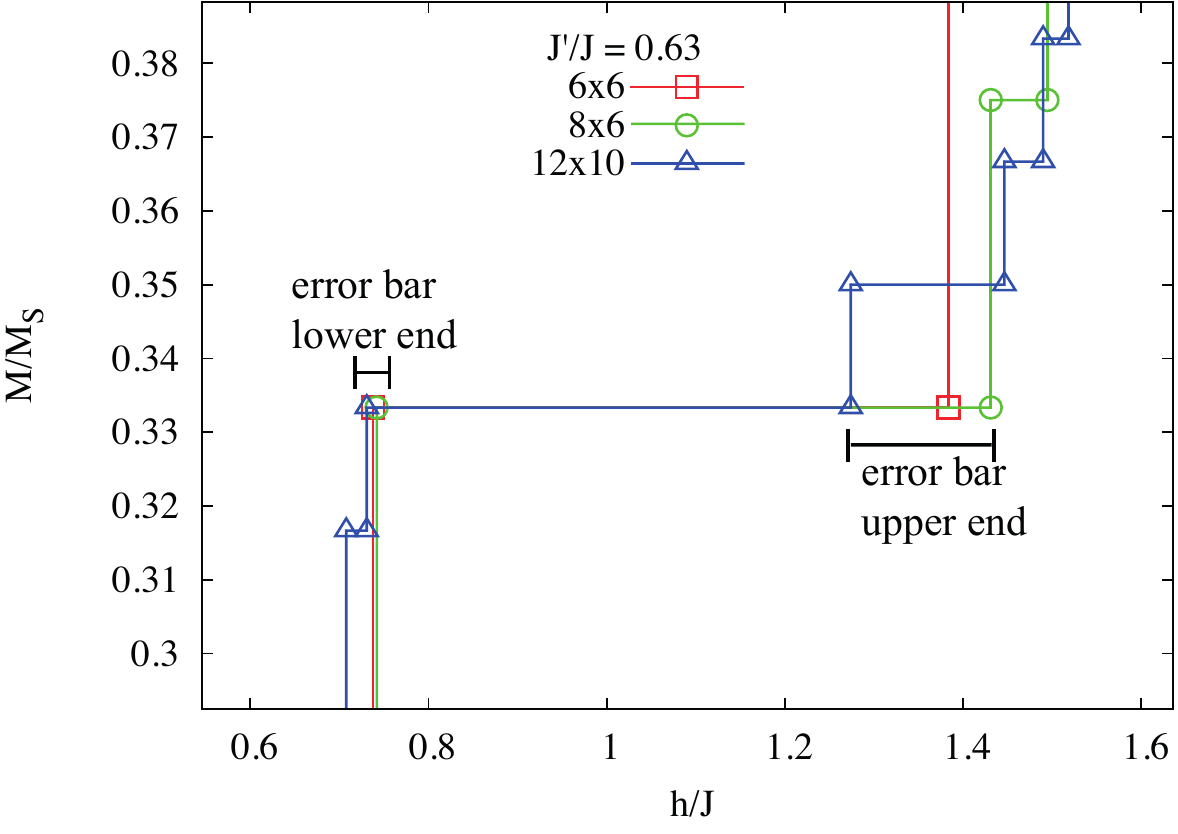}
\caption{ 
(a) 'Raw' DMRG results for the magnetization curves obtained for the different system sizes at $J'/J = 0.63$.  
(b) Zoom into the region containing the 1/3 plateau with data for the system sizes allowing for a plateau at this value of $M$. We estimate the error bars for the positions of the plateaus by considering the values for the different system sizes; as can be seen, for the lower end, a rather small error bar is obtained, while at the upper end the error bar is significantly larger, possibly indicating the presence of a shallowly growing magnetization curve following this plateau. 
}
\label{fig:DMRG}
\end{figure*}

For the results presented here, typically we perform 40 sweeps and keep up to $m=2000$ density matrix eigenstates. 
The resulting discarded weight is typically of the order of $10^{-4}$ or smaller. 
Despite the large number of sweeps and states kept, the energy can still change in the course of the last DMRG sweeps, so that in estimating the error bar additional caution needs to be taken. 
We obtain the magnetization curves shown in Fig.~\ref{fig:DMRG}(a). 
(The results for systems with more than $10\times8$ spins are obtained from comparing runs with different DMRG parameters and by taking the lowest achieved energies for a given value of $S^z_{\rm total}$). 
The sizes and positions of the plateaux at 1/4, 1/3, and 1/2 agree well with the iPEPS results. 
As can be seen, the data seems to collapse to a continuous magnetization curve in the high-field regions, but the accuracy is not high enough to exclude further plateaux at 2/3 and 3/4, and possibly additional values of $M$.  
In the low field region, the lower the magnetization and the larger the system size, the more difficult it is to reach convergence, so that DMRG data below $M=1/4$ needs to be considered with care; also, for the larger system sizes, it is difficult to obtain a unifying picture from the results between $M=1/4$ and $M=1/2$. 

In Fig.~\ref{fig:DMRG}(b) we show at the example of the $M=1/3$ plateau how we estimate the error bars in the extensions of the plateaus by comparing the results for the critical fields for the different system sizes. 
In Fig.~\ref{fig:mcurves}(b) below we compare the resulting endpoints of the plateaus 
to the ones obtained by iPEPS for $J'/J = 0.63$ (see next section). 
Good agreement is obtained; however, at the upper end of the 1/3 plateau the error bars are large. This might indicate a shallow increase of the magnetization at the end of this plateau, which would make it difficult to obtain the end point with a high accuracy.

\subsection{Infinite projected entangled-pair states}
\label{sec:iPEPS}
An infinite projected entangled-pair state (iPEPS) is an efficient variational
ansatz for a wave function in two dimensions in the thermodynamic
limit~\cite{verstraete2004,jordan2008,corboz2010}. It can be seen as a
two-dimensional extension of a matrix product state (MPS) -- the underlying variational ansatz of the density-matrix renormalization group method.
An iPEPS on the square lattice consists of a unit cell of 5th order tensors
which is periodically repeated on the lattice \cite{corboz2011}.
Each tensor has one physical index, which runs over the $d$ basis states of a lattice site, and four auxiliary indices with a certain bond dimension $D$  which connect to the four nearest neighboring tensors. The number of variational parameters per tensor is $d D^4$, thus the larger $D$ the (potentially) more accurate the ansatz. 

As a local basis we take the four basis states of a dimer, i.e., we simulate a square lattice model where each lattice site corresponds to one of the orthogonal dimers. For details on the simulations and iPEPS we refer to Ref.~\onlinecite{corboz2013} where a similar simulation setup was used for the Shastry-Sutherland model without an external magnetic field. The results presented are obtained with the so-called simple update in iPEPS, which gives a reasonably good estimate for the energy, and we checked several simulations with the more accurate (but computationally more expensive) full update (see Ref.~\onlinecite{corboz2010} for details). 

By using different unit cell sizes an iPEPS can represent different
translational symmetry broken states. To find the ground state for each value of
$H$ and $J'/J$ we have performed simulations with various rectangular unit cell
sizes up to $18\times 18$ to determine which cell yields the lowest variational
energy. We have run simulations up to $D=6$ for the supersolid phases (up to
$D=8$ for $J'/J=0.63$), and up to $D=8$ for the states within a plateau where we
exploited the (unbroken) U(1) symmetry \cite{bauer2011}.

To obtain an estimate of the energy in the infinite $D$ limit we linearly
extrapolate the finite $D$ data in $1/D$, which gives a value $E_\text{extrap}$.
Empirically we find that the energy converges faster than linearly in $1/D$, 
thus $E_\text{extrap}$ is likely to underestimate the true energy. As an
estimate we therefore take the mean between this value and the value at the
largest $D$, i.e., $E_{D=\infty} = (E_{D_{\text{max}}} + E_\text{extrap} )/2$,
and a rough estimate of the error bar is provided by half of the difference
between these two values, i.e., $\Delta =(E_{D_{\text{max}}} - E_\text{extrap} )/2$.

We determine the phase transition between two phases by determining the intersection point of the energies $E_{D=\infty}$  of the two adjacent phases, as e.g. done in Refs.~\onlinecite{corboz2010,corboz2013,messio13}. To obtain an estimate of the error bar on the phase transition we determine the intersection of the energies, where we take a lower (underestimated) value for the energy in one of the phases, $E_{\text{low}}=E_{D=\infty}-\Delta/2$, and a higher (overestimated) value of the energy in the other phase $E_{\text{high}}=E_{D=\infty}+\Delta/2$. This will lead to a shift of the phase boundary towards the second phase. Similarly, we take $E_{\text{high}}$ for the first phase and intersect it with $E_{\text{low}}$ of the second phase to obtain the other part of the error bar.

For a transition between a plateau phase and an adjacent supersolid phase we
find that the phase boundary moves towards the plateau phase with increasing
$D$, i.e., at finite values of $D$ the size of a plateau is overestimated. Therefore, we can obtain an upper bound of the phase boundary by taking the intersection of the energies of the two phases at a fixed $D=6$. 

The transition between the 1/3 plateau state and the 1/3 supersolid phase is found to be of second order. In order to have a lower bound on the phase boundary we linearly extrapolate the magnetization $M(H)$ and intersect it with $M=1/3$. (We checked for $J'/J=0.63$ and full update iPEPS simulations that this extrapolation yields a lower bound on the phase boundary). We use this estimate of the lower bound also for the transition between the 2/5 plateau and 2/5 supersolid, and the 1/2 plateau and the 1/2 supersolid phase.

\subsection{Series expansion}
In the following we shortly explain how we have implemented high-order series
expansions for the magnetization plateaux at $M=1/2$, $M=2/5$, $M=1/3$, and
$M=1/4$. For each plateau we calculated the ground-state energy per dimer.
Additionally, the one-particle gap for the plateau structures at $M=1/2$ and
$M=1/3$ is determined.

\begin{figure}[tb!]
\begin{center}
\parbox{16.3cm}{\includegraphics[width=16.1cm]{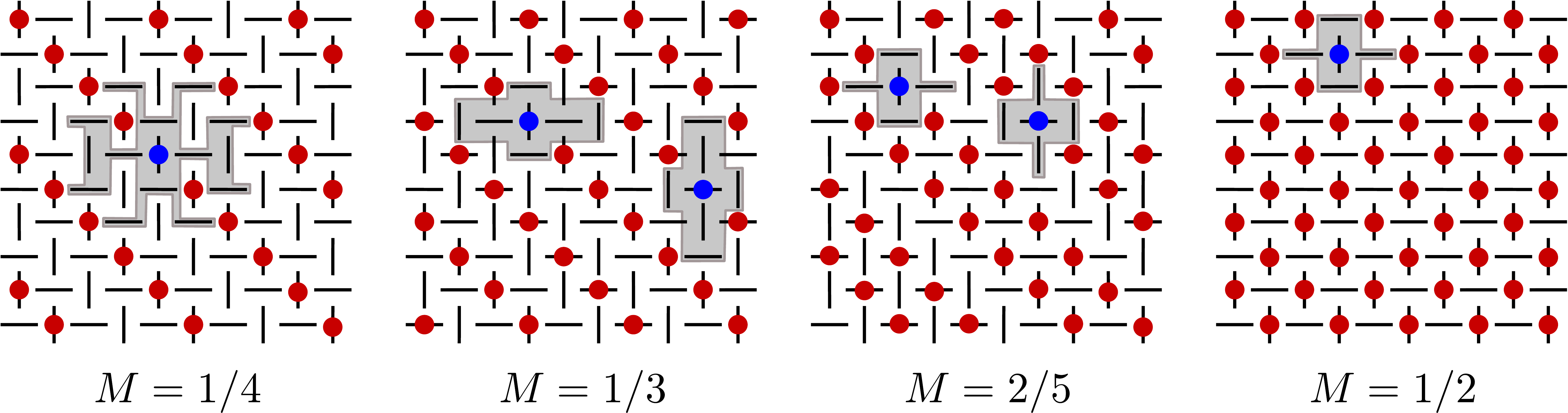}}
\caption{Illustration of the plateau structures considered by series expansion. Red dots denote triplets $t^1$ while empty dimers correspond to the presence of singlets $s$. The gray areas $\Gamma^M_i$ contain for each plateau all repulsive interactions involving the blue triplet on dimer $i$ which have to be put in the unperturbed Hamiltonian $\hat{H}_0$.}
\label{fig:structures}
\end{center}
\end{figure}

We aim at setting up a high-order series expansion for the most relevant
magnetization plateaux at $M=1/2$, $M=2/5$, $M=1/3$, and $M=1/4$. The idea is to {\it deform} the Shastry-Sutherland model such that one can define an unperturbed Hamiltonian $\hat{H}_0$ which has as a unique ground state a product state of singlets $s$ and triplets $t^1$ with the desired magnetization $M$ and plateau structure as illustrated in Fig.~\ref{fig:structures}.   

Physically, this is achieved in two steps. First, we add a magnetic field of strength $J$ to $\hat{H}_0$. As a consequence, on an isolated dimer one has two degenerate low-energy states, the singlet $s$ and the triplet $t^1$, while the other two states $t^0$ and $t^{-1}$ cost a finite energy. Second, one adds repulsive interactions between triplets $t^1$ to $\hat{H}_0$ and by subtracting the same kind of interactions in the perturbation $\hat{V}$ introducing the perturbative parameter $x\in\left[ 0,1\right]$. The Shastry-Sutherland model is then recovered for $x=1$. To be specific, we define 

\begin{eqnarray}
\label{eqn:H}
{\hat H}_{M} &=& J\left( \hat{H}_{J} +  \hat{H}_{h} +  \lambda\hat{H}_{M}\right) + x\left( J' \hat{H}_{J'} -  \lambda J \hat{H}_{M}\right) + (h-J) \hat{H}_{h},\\
              &=& \hat{H}_0 + x \hat{V}+ (h-J) \hat{H}_{h}\quad ,
\end{eqnarray}
where $\hat{H}_{M}=\sum_i\hat{n}_i \sum_{j\in\Gamma^M_i} \hat{n}_j$ with $\hat{n}_i=\hat{t}^\dagger_{1,i}\hat{t}^{\phantom{\dagger}}_{1,i}$. Here $\Gamma^M_i$ corresponds to a specific collection of dimers around dimer $i$ which can differ for each plateau structure. The $\Gamma^M_i$ are illustrated in Fig.~\ref{fig:structures} as gray areas. The parameter $\lambda$ is a parameter one can choose freely which might result in an improved convergence of the series. Here we have chosen $\lambda=1$ for all plateaux except $M=1/2$ where a $\lambda<1$ gives better results.  

The series expansion is done in the perturbative parameter $x$. We used
Loewdin's projector method \cite{Loewdin62} to calculate the ground-state energy
per dimer $\epsilon_M$ in the thermodynamic limit. We have obtained order 9 for
$\epsilon_M$ with $M=1/2$, $M=2/5$, $M=1/3$ and order 8 for $\epsilon_{1/4}$.
Additionally, we used Takahashi's degenerate perturbation theory
\cite{Takahashi77,Klagges12} to calculate the one-particle gap $\Delta_M$ for
$M=1/2$ and $M=1/3$. Here we have calculated order 9 for $M=1/2$ and order 7 for
$M=1/3$ \footnote{We also have calculated the one-hole gap for $M=1/2$ and $M=1/3$ (removing one triplet $t^1$ from the plateau structure), but it plays no role for the magnetization curve of the Shastry-Sutherland model.}. In all cases one has to fix the ratio $J'/J$ and one has to perform the extrapolation in $x$ up to $x=1$.

One can deduce two kinds of information from the different series expansions: i) Location of first-order phase transitions between two different plateaux. To this end one defines $f_M(h)=\epsilon_M+(h-J)M$. A first-order phase transition between the plateaux with magnetizations $M$ and $M'$ then takes place at $h_{\rm 1st}$ for which $f_M(h_{\rm 1st})=f_{M'}(h_{\rm 1st})$. ii) The breakdown of a plateau with magnetization $M$ by a second-order phase transition can be located by the help of the one-particle gap $\Delta_M$. If one finds $\Delta_M<h_{\rm 1st}$ for a fixed ratio $J'/J$, then one expects a second-order phase transition at $h=\Delta_M+J$ to a supersolid phase with the same kind of crystalline order. Note that the series expansion is not sensitive to first-order phase transitions to other plateaux with different $M$.

\begin{figure}[tb!]
\begin{center}
\includegraphics[width=17cm]{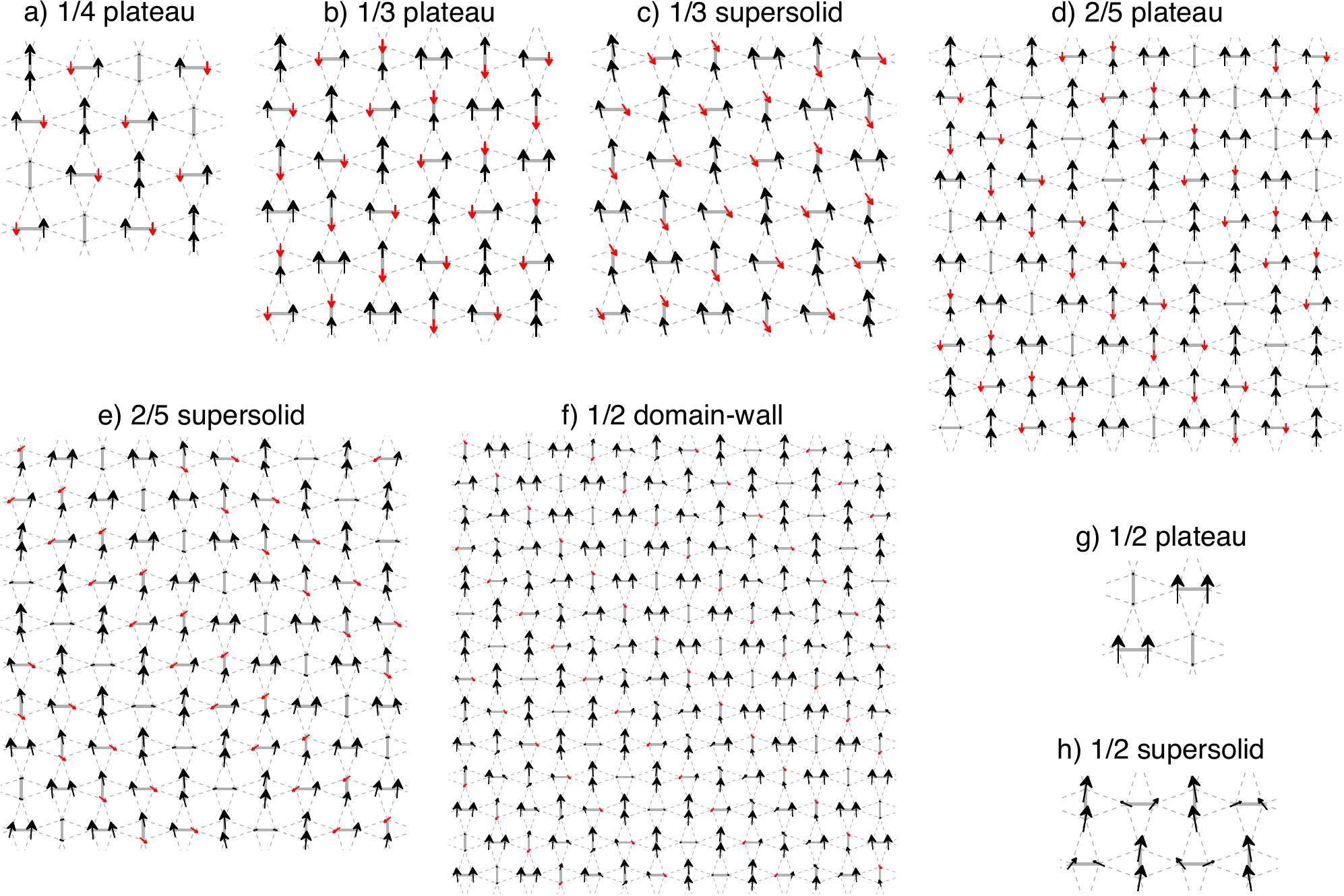}
\caption{Spin structure of the different phases obtained in different rectangular unit cells with iPEPS. The field direction is along the vertical axis, where the black spins predominantly point along the field, and the red spins predominantly point in the opposite direction of the field. The states in the plateaux have a vanishing transverse component, in contrast to the supersolid phases. Each supersolid state above a certain plateau is obtained by deforming (rotating) the spins of the state in the plateau. The 1/2 domain-wall phase contains stripes of the 1/2 plateau state, separated by (superfluid) domain walls. 
}
\label{fig:states}
\end{center}
\end{figure}

\section{Overview of phases}
\label{sec:phases}
In Fig.~\ref{fig:states} we present the spin structures of the different phases mentioned in the main text. These phases have been obtained with iPEPS using different rectangular unit cells, as explained in Sec.~\ref{sec:iPEPS}.  

We note that slightly above the 1/2 plateau state there is also a competing 2/5 supersolid phase. However, we have found that the 1/3 supersolid phase is energetically slightly lower.


\section{Comparison between numerical methods and experimental data}
\label{sec:comp}

\subsection{Extent of the 1/3 and the 1/2 plateaux}
\begin{figure}[tb!]
\begin{center}
\includegraphics[width=8.5cm]{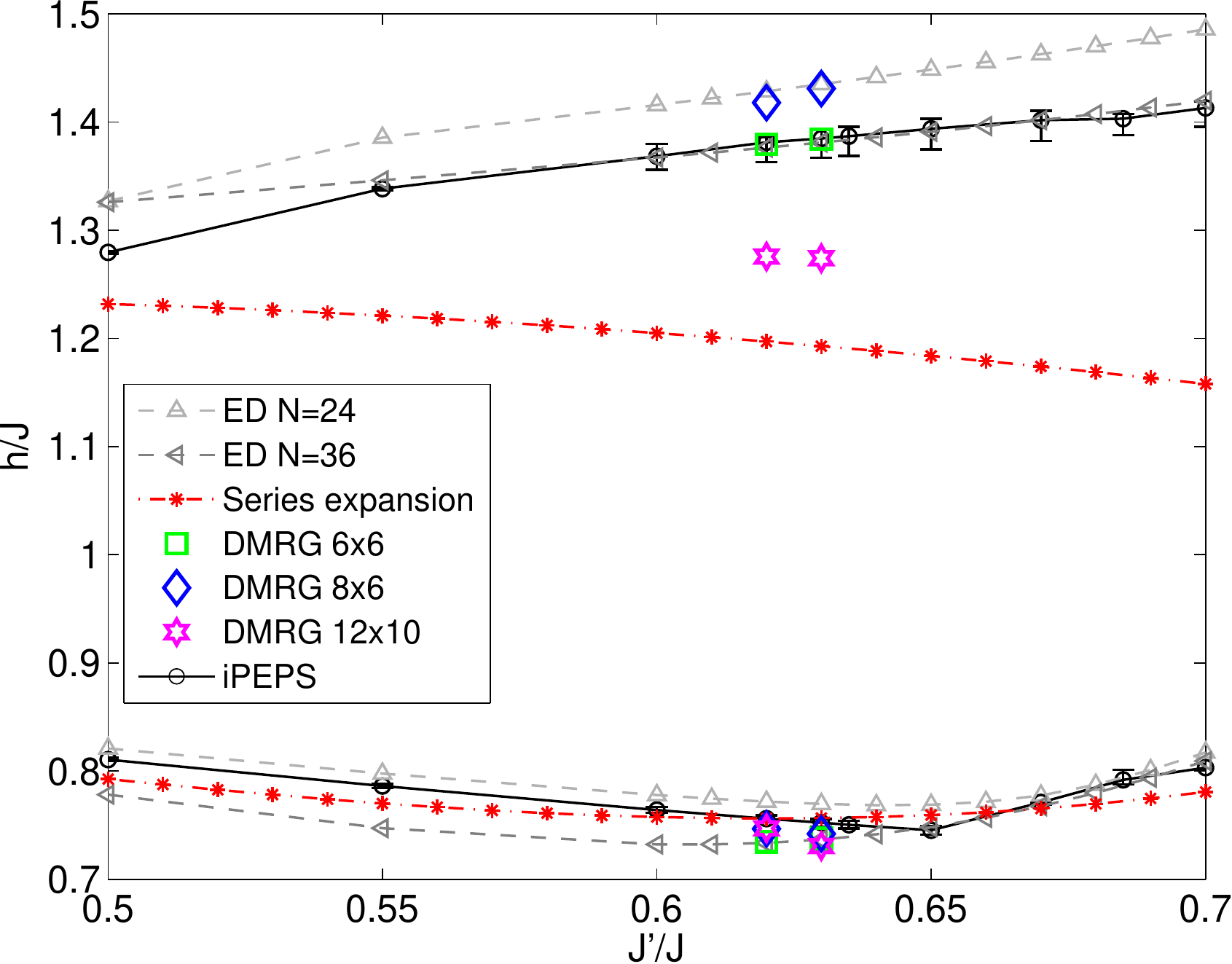}
\includegraphics[width=8.5cm]{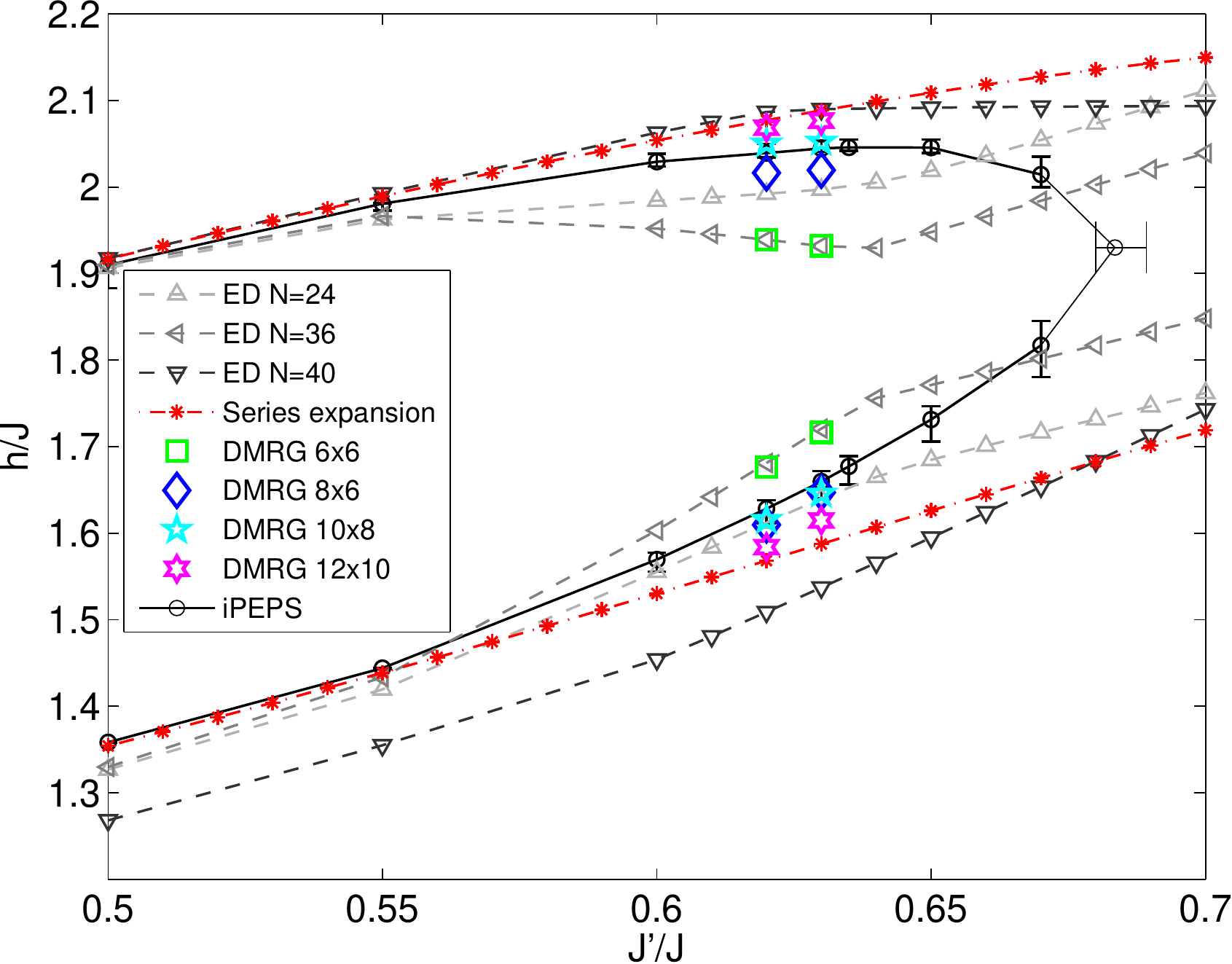}
\caption{Upper and lower boundary of the 1/3 plateau (a) and the 1/2 plateau (b) obtained with different methods. }
\label{fig:plateaux}
\end{center}
\end{figure}

In Figure~\ref{fig:plateaux} we compare the numerical results of the phase boundaries of the 1/3 plateau (a) and the 1/2 plateau (b) as a function of $J'/J$, obtained with the various methods.

A good agreement between all the methods is found for the lower edge of the 1/3
plateau in the whole parameter range $[ 0.5, 0.7]$ considered here. The series
expansion results lie close to the iPEPS values. The $N=36$ lattice
used in ED (and DMRG) is compatible with the structure of the 1/3 plateau state, but not
with the 1/4 plateau state. This explains why the $N=36$ lattice overestimates 
the extent of the 1/3 plateau on the lower edge for $J'/J \lesssim 0.65$, where we
find a transition between the 1/4 plateau and the 1/3 plateau. For larger $J'/J$
iPEPS predicts a transition between a $1/3$ supersolid phase and the $1/3$
plateau, i.e., structures which are both compatible with the $N=36$ lattice, and therefore the agreement is better. 

For the upper edge of the 1/3 plateau a good agreement between iPEPS, and the 
$N=36$ lattice from ED and DMRG can be found. This lattice is compatible with the structures of
both adjacent phases, except for $J'/J \lesssim0.55$ where iPEPS and SE predict a
direct transition between the 1/3 plateau and the 2/5 plateau. The latter is not
compatible with the $N=36$ lattice and this is why the extent of the 1/3 plateau is overestimated for $J'/J \lesssim0.55$. We find a large deviation between SE and the other methods, which is difficult to explain. One possibility is that the phase transition is of second order with a very slow increase of the order parameter as a function of $J'/J$, which would be difficult to capture with the other methods. The low transition value found with DMRG on the $12\times 10$ system also points towards this possibility.   From iPEPS, however, we do not find indications for such a small order parameter over a wide range of $J'/J$. Nevertheless, such a scenario would still be compatible with the experimental data (and it could explain the slow increase of $M/M_S$ in the 1/3 plateau). 

In any case, the 1/3 plateau is rather wide in the parameter regime under
consideration. This might be attributed to the proximity to a classical plateau
state at $J'/J=1/2$ \cite{Moliner09}.

For the lower edge of the 1/2 plateau we find a good agreement between iPEPS and
SE for $J'/J \lesssim0.55$, where both methods predict a direct transition between
the 2/5 plateau and the 1/2 plateau. However, for $J'/J \gtrsim0.55$ iPEPS predicts
a supersolid (or domain-wall) phase adjacent to the 1/2 plateau, which are not
captured in the SE calculations, and this leads to an overestimation of the
extent of the 1/2 plateau with SE.  There are rather large variations of the
phase boundary for the different ED lattices. The iPEPS phase boundary lies in
between the $N=24$ lattice and the $N=36$, $N=40$ lattices (for $0.55<J'/J <0.65$).

Also for the upper edge of the 1/2 plateau SE agrees with the iPEPS result for
$J'/J \lesssim0.55$, where both methods predict a transition between the 1/2 plateau
and the 1/2 supersolid. For larger $J'/J$ iPEPS finds a 1/3 supersolid with a lower
variational energy than the 1/2 supersolid, which explains the deviation from
the SE phase boundary. Large finite-size effects are found with ED also for the
upper edge. The iPEPS result lies in between the $N=24, 36$ and the $N=40$ phase
boundary for $J'/J <0.65$. The 1/2 plateau obtained with ED does not close in
the considered parameter range due to finite-size effects. 

Finally, the iPEPS results for the extent of the 1/2 plateau are also compatible with the finite size DMRG data, where the best agreement is found with the $10\times8$ system.

\begin{figure}[tb!]
\begin{center}
\includegraphics[width=8.5cm]{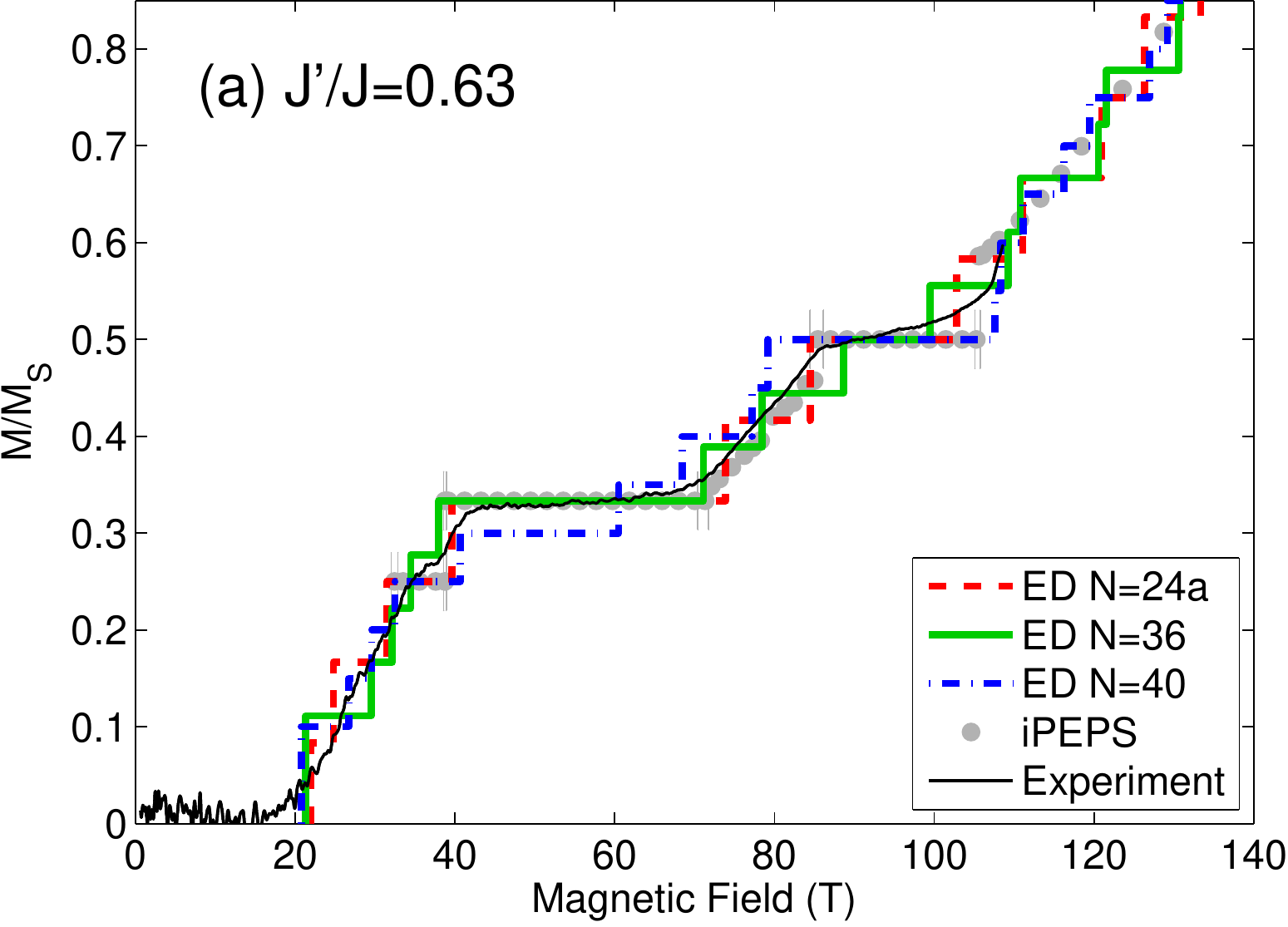}
\includegraphics[width=8.5cm]{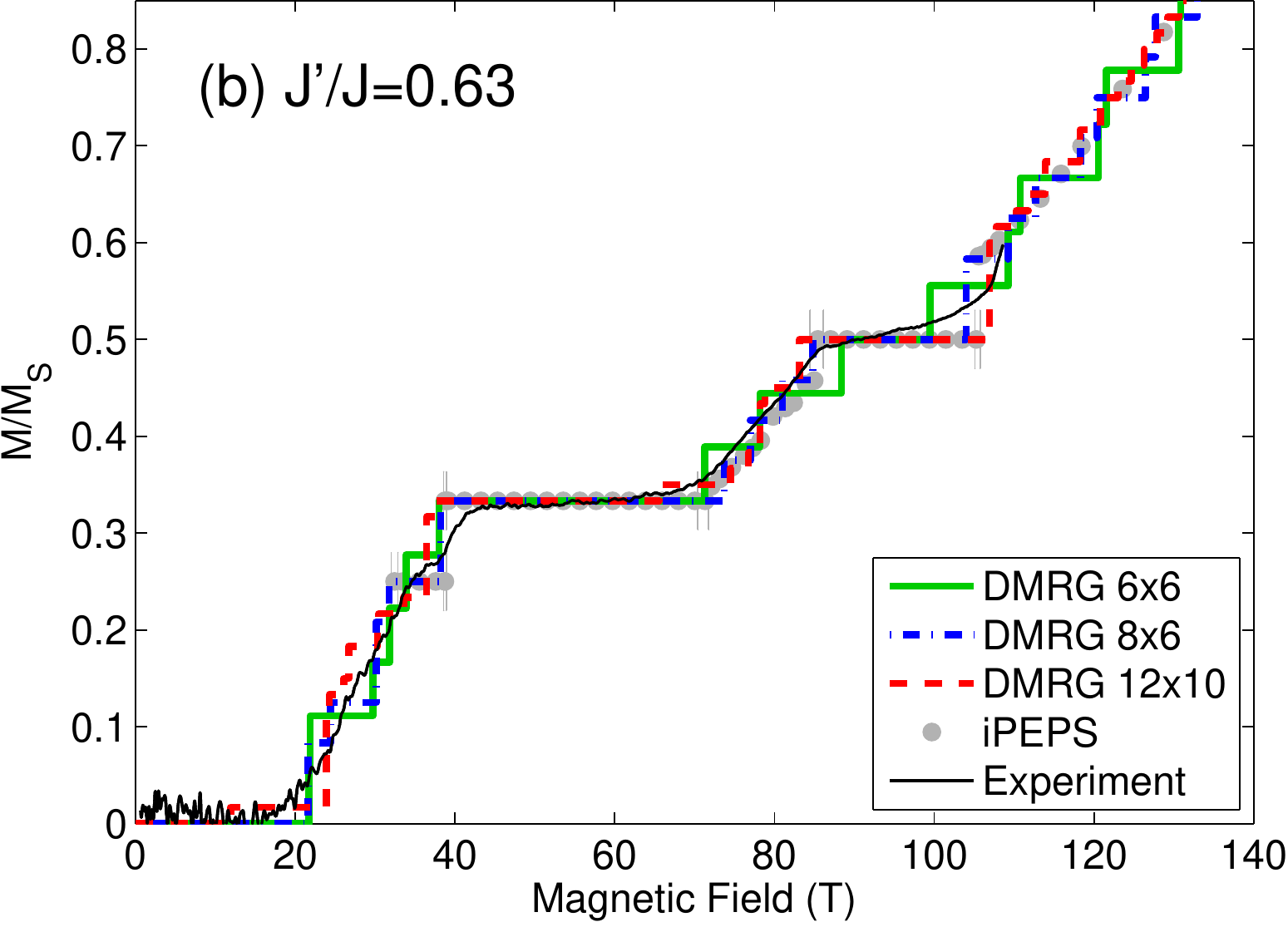} \vspace{0.2cm}

\includegraphics[width=8.5cm]{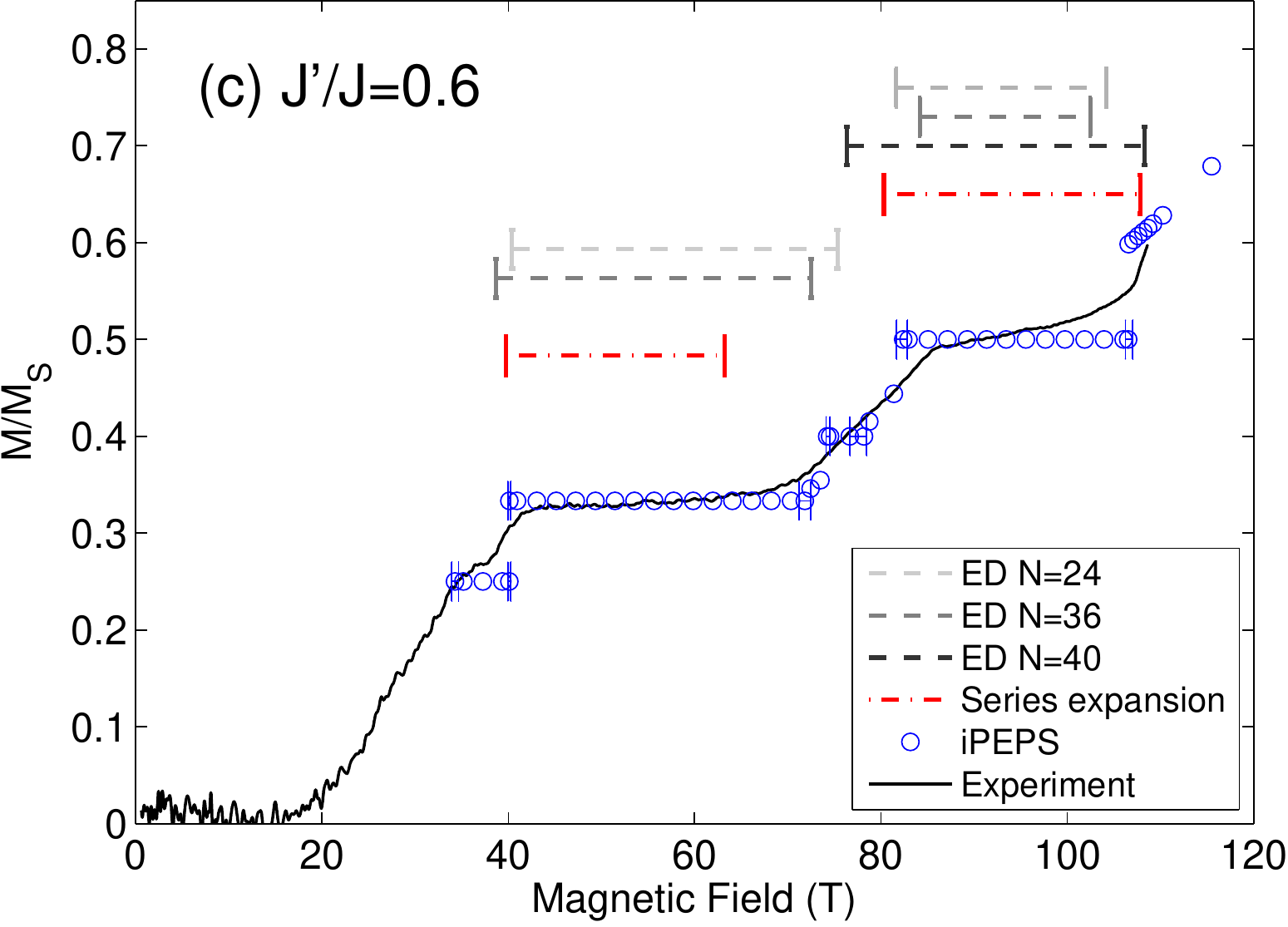}
\includegraphics[width=8.5cm]{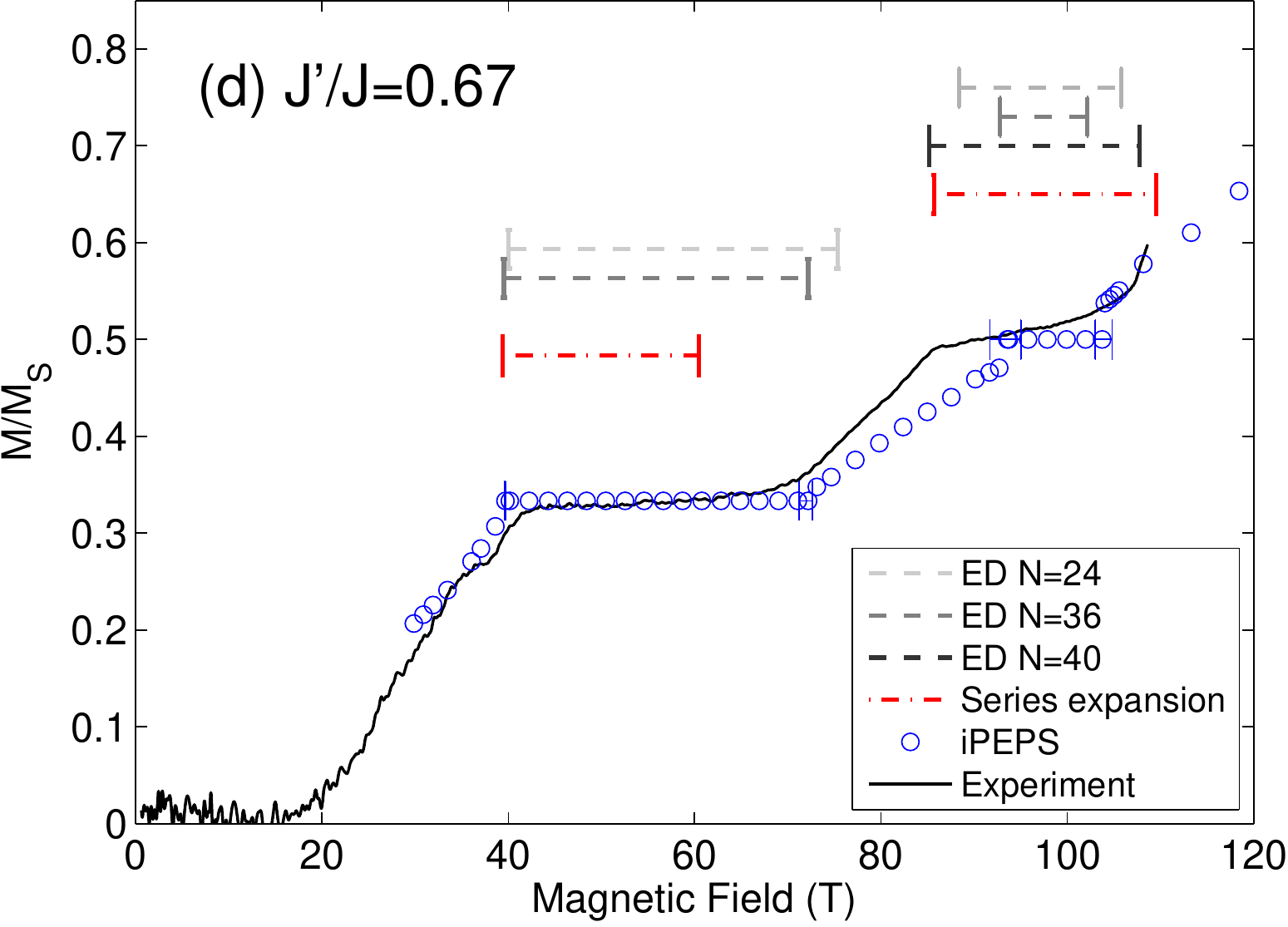}
\caption{Magnetization curves for $J'/J=0.63$ obtained from experiments and
iPEPS, compared to (a) ED, and (b) DMRG. (c-d) The experimental data is not
compatible with the numerical magnetization curves for  $J'/J=0.6$ (c) and $J'/J=0.67$ (d).}
\label{fig:mcurves}
\end{center}
\end{figure}

\subsection{Comparison of magnetization curves}
In Fig.~\ref{fig:mcurves} we present a comparison of the magnetization curves between the different methods and the experiment.

Figure~\ref{fig:mcurves}(a) shows a plot obtained with ED for different system
sizes for $J'/J=0.63$. Variations in the magnetization curves can be found for
different lattice sizes, but there is an overall good agreement with iPEPS and the experimental data.
 
A good agreement is also found with DMRG as shown Fig.~\ref{fig:mcurves}(b), although the finite-size effects on the upper edge of the 1/3 plateau are rather large.

In Fig.~\ref{fig:mcurves}(c) we present an attempted fit between iPEPS and experimental data for $J'/J=0.6$, showing several mismatches. The 1/2 plateau is considerably bigger than in the experimental data, and we clearly find a 2/5 plateau at this value for $J'/J=0.6$, which is absent in the experiment. We therefore conclude that $J'/J=0.6$ is too small. 

A bad fit is also obtained if  $J'/J$ is too large, as shown in Fig.~\ref{fig:mcurves}(d) for $J'/J=0.67$. The 1/2 plateau turns out to be too small in this case, and the slope of the magnetization curve between the 1/3 and 1/2 plateau is not as steep as in the experiment.

\bibliography{SCBO}